%% file: ms.tex
\documentclass[12pt, preprint]{aastex}

\def\cc{cm$^{-3}$}
\def\kms{km s$^{-1}$}

\def\ammonia{NH$_3$}
\def\h2{H$_2$}
\def\n2h{N$_2$H$^+$}
\def\Ms{M$_\odot$}

\def\13co{$^{13}$CO}
\def\c18o{C$^{18}$O}
\def\nh{$n(H_2$)}
\def\lp{\>\> .}
\def\lc{\>\> ,}
\def\cm2{cm$^{-2}$}
\def\nh{n(H$_2$)}

\def\Ms{$M_{\odot}$}
\def\micron{$\mu$m}

\def\h2{H$_2$}
\begin{document}

\title{Massive Quiescent Cores in Orion. -- II. Core Mass Function}
\author{D. Li\altaffilmark{1}, T. Velusamy\altaffilmark{1}, P. F. Goldsmith\altaffilmark{1},  
	and W. D. Langer\altaffilmark{1} }
\affil{Jet Propulsion Laboratory, California Institute of Technology}
\altaffiltext{1}{JPL/Caltech, 4800 Oak Grove Drive,  Pasadena, CA 91109.  Please send preprint request to dili@jpl.nasa.gov.}

%**********************************************************************************
\begin{abstract}
We have surveyed submillimeter continuum emission from relatively quiescent regions  in the Orion molecular cloud to determine how the core mass function in a high mass star
forming region compares to the stellar initial mass function.  
Such studies are important for understanding the evolution of cores to stars, and for comparison to formation processes in high and low mass star forming regions.
We used the SHARC II camera on the Caltech Submillimeter Observatory telescope to obtain 350 \micron\  data having 
angular resolution of about 9 arcsec, which corresponds to 0.02 pc at the distance of Orion. 
Further data processing using a deconvolution routine enhances the resolution to about 3 arcsec.
Such high angular resolution allows a rare look into individually resolved dense structures in a massive star forming region. 

Our analysis combining dust continuum and spectral line data defines a sample of 51 Orion molecular cores with masses ranging from 
0.1 \Ms\  to 46 \Ms\  and a mean mass of 9.8 \Ms, which is one order of magnitude higher than the value found in typical low mass star forming regions, such as Taurus. 
The majority of these cores cannot be supported by thermal pressure
or turbulence, and are probably supercritical.  
They are thus likely precursors of protostars.

The core mass function for the Orion quiescent cores can be fitted by a power law with an index equal to -0.85$\pm$0.21.   This is significantly flatter than the Salpeter initial mass function and  is also flatter than the core mass function found in low and intermediate star forming regions. 
When compared with other massive star forming regions such as NGC 7538, 
this slope is flatter than the index derived for samples of cores with masses up to thousands of \Ms.
Closer inspection, however, indicates slopes in those regions similar to our result if only cores in a similar mass range are considered.
Based on the comparison between the mass function of the Orion quiescent cores and those of cores in other regions, we find that the core mass function is flatter in an environment affected by ongoing high mass star formation. 
Thus, it is likely that environmental processes play a role in shaping the stellar IMF later in the evolution of dense cores and the formation of stars in such regions.

\end{abstract}
\setcounter{footnote}{0}

%*******************************************************************************
\section{INTRODUCTION}

The formation processes of high mass and low mass stars have long been suggested to be different.
High mass  star formation (HMSF) may require supercritical conditions in the 
natal clouds, while low mass star formation (LMSF) mostly occurs 
in subcritical gas (e.g.\ Shu, Adams \& Lizano 1987). Massive stars may also be formed through
stellar mergers (Bonnell \& Bate 2002).
The recent reports of  disks around massive stars (Chini et al.\ 2004; Patel et al.\ 2005), 
however, suggests that massive stars can be formed through disk accretion, the same way as low mass stars.
For better understanding massive star formation, 
it is important to obtain the physical conditions of HMSF regions.

Past observations of molecular clouds that 
harbor young stars of different masses find  
clear differences between HMSF regions and LMSF regions.
High mass stars form only in GMCs while low mass stars can form in dark clouds as well as in GMCs. 
High mass stars are predominantly formed in clusters, while low mass stars may form in isolation. 
The star formation efficiency is generally higher in HMSF regions (e.g.\ Myers et al. 1986 and Lada \& Lada 2003). 
The gas properties in these regions also differ. 
In HMSF regions, the density and temperature tend to be higher, and spectral line widths greater indicating higher degree
of turbulence. 
The causal relationship between the presence of young high mass stars, different molecular cloud characteristics, 
and the possible variation of the star formation processes, however, is not well established. 

The objects directly connecting general molecular cloud material and young stars are
the so called 'cores', which are condensations with elevated density and extinction 
and are likely to be bound by gravity (Ward-Thompson et al.\ 2006). Cores are potential 
precursors of protostars.
The density and temperature structure of quiescent cores (no  
IRAS point sources and no association with molecular outflows) 
 provides important constraints for distinguishing between star formation models and determining the 
initial conditions of star formation.  
For example, although a singular isothermal sphere leads to an inside--out collapse 
at constant accretion rate (Shu 1977), the evolution becomes quite 
different in the non--isothermal case (Foster \& Chevalier 1993), or in situations 
involving nonthermal pressure support in the form of turbulence or magnetic fields 
(see, for example, Ward-Thompson, Motte \& Andre 1999).  
Each of these models makes different predictions concerning the form of the 
radial density profiles in cold prestellar cores and their evolutionary successors, presumed to be represented by Class 0 protostars. 
The need to test these models has motivated extensive studies of the density profiles of prestellar cores using far-infrared and submillimeter imaging observations (see for example, Evans et al.\ 2001).

Previous detailed studies of dense cores have largely been focused on low mass star forming regions, such as Taurus and Ophiuchus. 
These regions are within 150 pc of the earth thus allowing high spatial resolution and strong signals.  
One important observational aspect to consider when tackling this problem is the mass 
function of dense molecular cores, especially in their relatively quiescent stages before 
being disrupted by the onset of stellar energy input.
The core mass function may bear clues to the relative universality of initial 
mass function (IMF) of stars, which remains a prominent question in the field of star formation.
If the core mass function resembles that of the stellar IMF, it is likely that the star formation 
processes within each core, including collapse, disk accretion, jet, and outflow, are uniform in 
terms of the efficiency of mass transfer from the ISM to stars. 
On the other hand, if the core mass function differs significantly 
from the stellar IMF, it is likely  that environmental factors, such as competitive accretion, 
shape the resulting IMF.  
Most of the past studies find a core mass function similar to the 
stellar IMF  (e.g. Motte, Andre \& Meri et al.\ 1998; Testi \& Sargent 1998; Young et al.\ 2006).  

Given the importance of quiescent cores in a massive star forming environment, we have 
aimed to obtain data for a sizable sample of such cores in Orion, which is the closest known GMC. 
At a distance of about 450 pc, it is possible to detect individual cores through submillimeter imaging.  
We have also obtained limited amount of spectral line data (Li et al.\ 2003, paper I), 
which help to constrain the temperature of these cores and their ambient environment. 
Adequate spatial resolution and knowledge of the velocity structure (thermal, turbulent, and systematic) are crucial for
obtaining an accurate estimate of the core mass and their dynamic state.
In this paper, we will focus on the determination of the mass function of these quiescent cores.

\section{Observations and Data Reduction}	 

\begin{figure}[htp]
   \includegraphics[scale=.70]{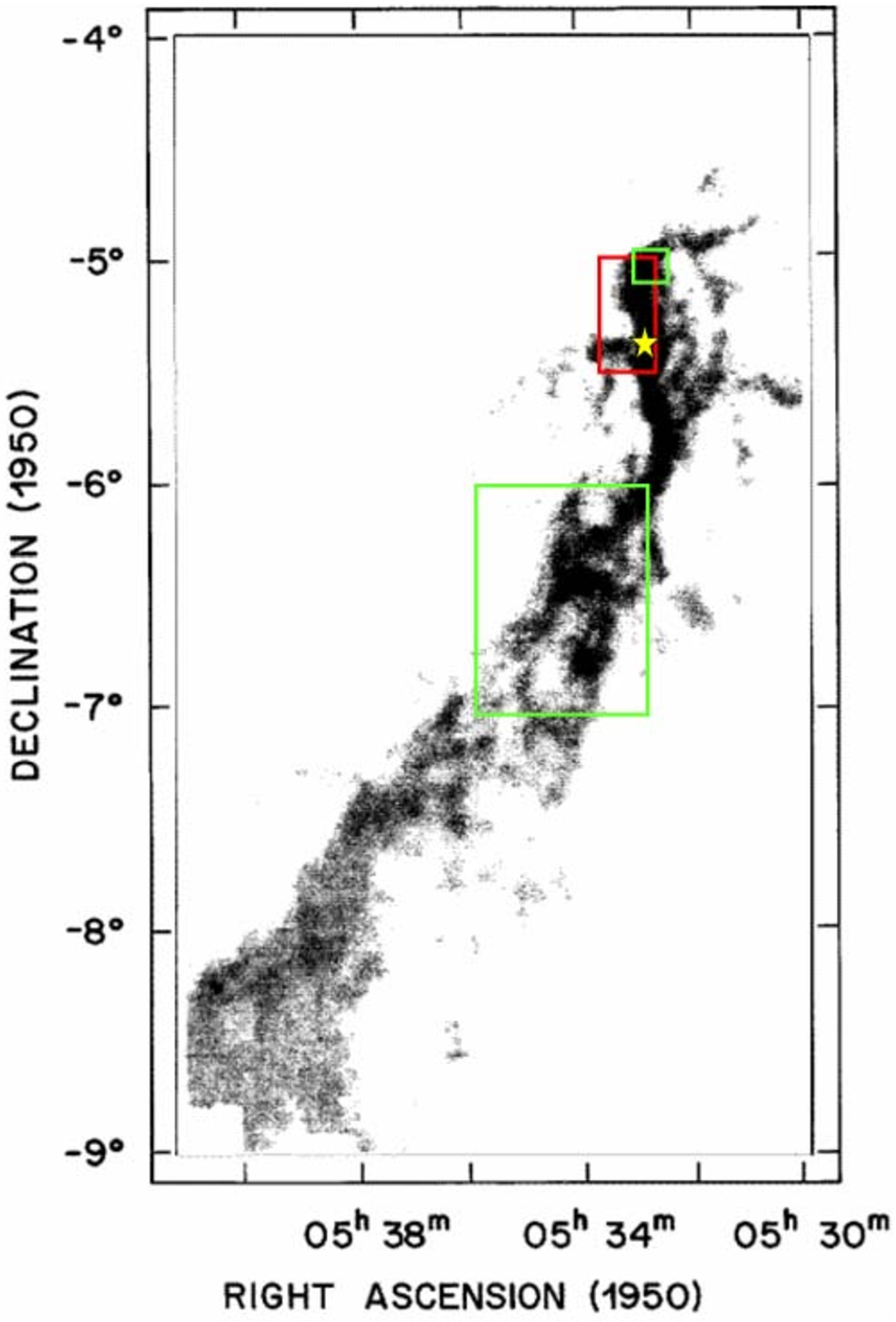}
    \caption{Our maps are in the green rectangles, which are overlaid on
a $^{13}$CO 1-0 integrated intensity map reproduced from Bally et al.\ (1987).
The red rectangle indicates, roughly, the coverage provided by the data from 
Lis et al.\ (1998). The yellow star indicates the location of the Trapezium cluster.} 
      \label{fig:coverage}
\end{figure}

We have used the Submillimeter High Angular Resolution Camera 
(SHARC II, see Dowell et al.\ 2003), 
installed on the 10.4 meter telescope of the Caltech Submillimeter Observatory (CSO) to carry out the survey of quiescent Orion cores. 
The fields which we have mapped are chosen based on the presence of dense gas and the 
absence of indications of active star formation, such as IRAS sources and outflows. 
A more detailed discussion of the selection criteria can be found in paper I. 
The locations of our maps are indicated in
Fig.~\ref{fig:coverage}.
SHARC II is a 12 by 32 bolometer array, and each pixel samples a region of 4.85\arcsec\ on the sky. 
The footprint size of the total array is 2.59\arcmin\ by 0.97\arcmin. 
The size of the individual maps in our data set is several times the footprint size, and the data were obtained by repeated scans of the telescope over the region of interest, using a scanning pattern (see below).

Orion is a region with bright, spatially extended emission at submillimeter wavelengths. 
The normal throw range of the chopping secondary is not enough for moving the reference beam completely off of the emission from the clouds. 
Chopping against the cloud emission background brings significant uncertainty
into the absolute calibration and decreases the capability of detecting sources of larger sizes (e.g. comparable to the chopper throw angle).  
Our maps are obtained in the non-chopping BOXSCAN mode of SHARC II. 
By scanning the desired region quickly and repeatedly in a complex pattern, each sky position is covered by every bolometer in the array and with 
different time variations. 
Such redundancy enables the sky image to be computed subsequently through software iteration. 
We have employed the Comprehensive Reduction Utility for SHARC II (CRUSH) developed by A.\ Kov{\'a}cs at Caltech.
This reconstruction has proven to be fairly stable and consistent in recovering both the bright peaks and extended structure in our maps.

During three observing runs from 2003 to 2005 at the CSO, we have obtained 350 \micron\ maps for 8 fields toward quiescent portions of Orion. 
The sizes of these fields ranges from 4'$\times$4' to 8'$\times$8'. 
About every 60 minutes, we obtained a short scan of a bright calibrator, such as Mars. 
The scans of the calibrator sources from each day are later reduced 
using the same CRUSH program to provide the absolute flux scale for our images. 
The CRUSH program corrects the data for atmospheric absorption utilizing measurements of
opacity ($\tau$) provided by a tipper operated at 350 \micron.
The fluxes of planets and point sources at the time of observation
are calculated by the FLUXES program obtained from JCMT.

\begin{figure}[htp]      	
   	\includegraphics[width=0.80\textwidth]{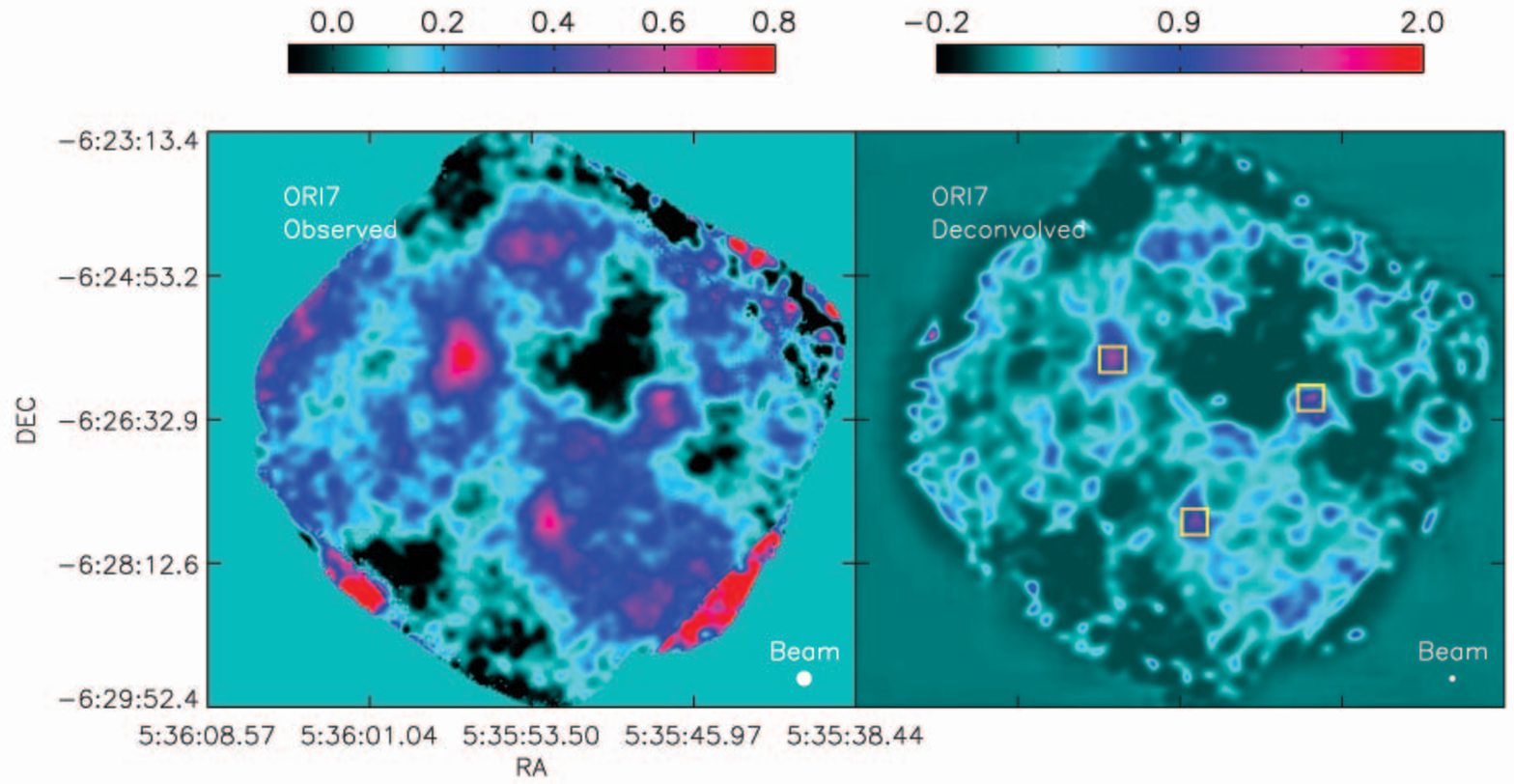}\\
   	\includegraphics[width=0.80\textwidth]{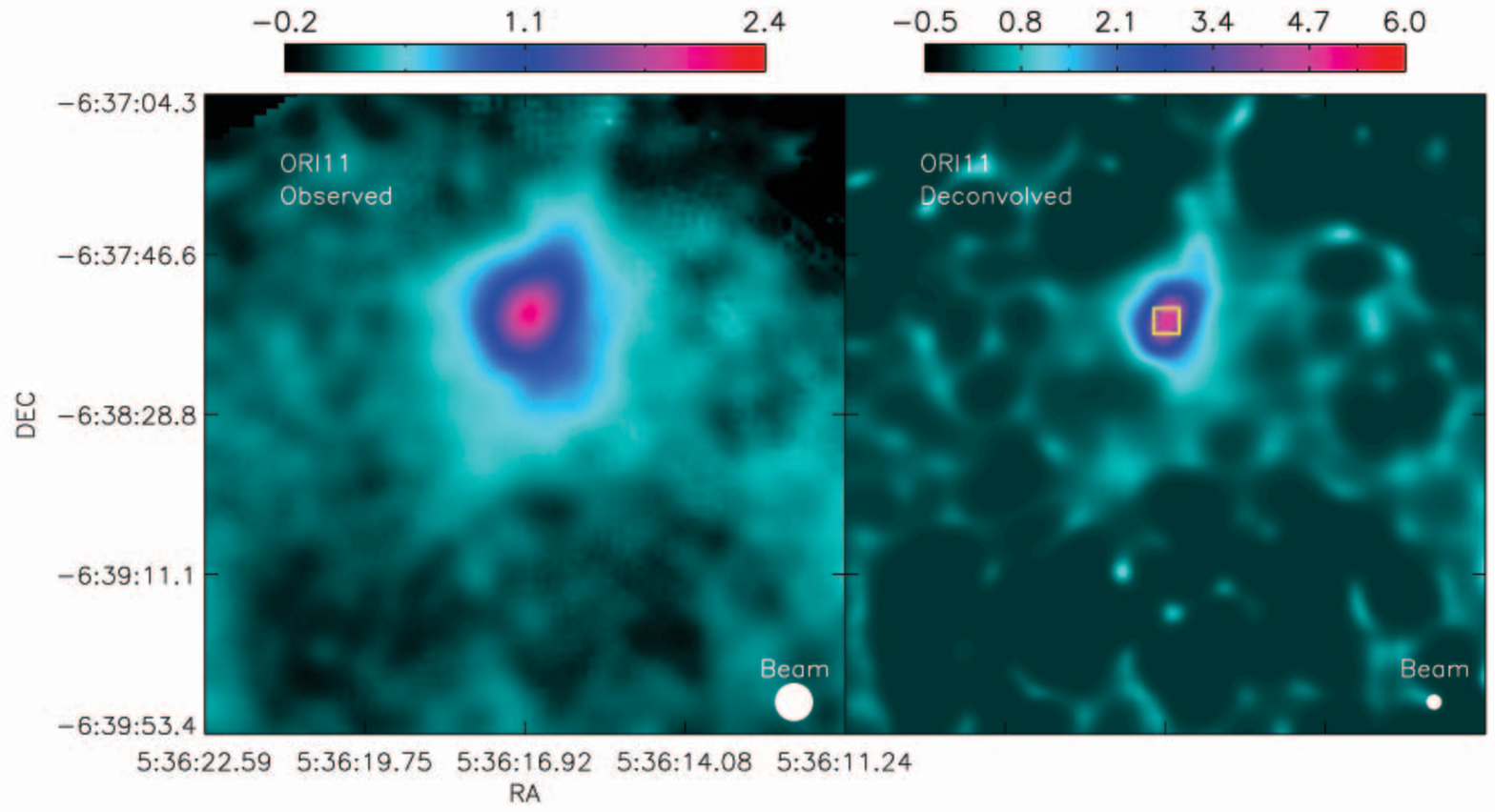}
   	\caption{350 \micron\ images of the Orion cores.  
The left hand panel is the observed image
 and the right hand panel is the deconvolved image. The coordinates are in J2000.  
Both images are plotted in units of Jy per 9\arcsec\ beam.  
The yellow squares indicate the peak positions of cores 
found by the Clumpfind program. The beam sizes of the observed and deconvolved images are shown as
filled circles. The upper panel shows the survey field ORI7, and the lower panel, ORI11. }
\label{fig:ori7-11}
\end{figure}

After image reconstruction by CRUSH and flux calibration, the typical RMS noise level on the background part of a image (i.e. devoid of cores with peak flux greater than about the 10 $\sigma$ level) is about 0.15 Jy/beam.
We further processed our calibrated data using the deconvolution program Hires (Backus et al.\ 2005) based on 
Richardson-Lucy iteration procedure (Richardson 1972; Lucy 1974).
The deconvolution procedure is based on an idealized beam pattern obtained from repeated observations of Neptune. 
The deconvolution processing is stable and generally converges within 50 iterations.
The edge pixels, where the signal to noise ratio is poor due to insufficient sampling, have been avoided during iteration (see the 
observed and deconvolved images of ORI7  in Fig.~\ref{fig:ori7-11}  for an example).

\begin{figure}[htp]
   	\includegraphics[width=0.80\textwidth]{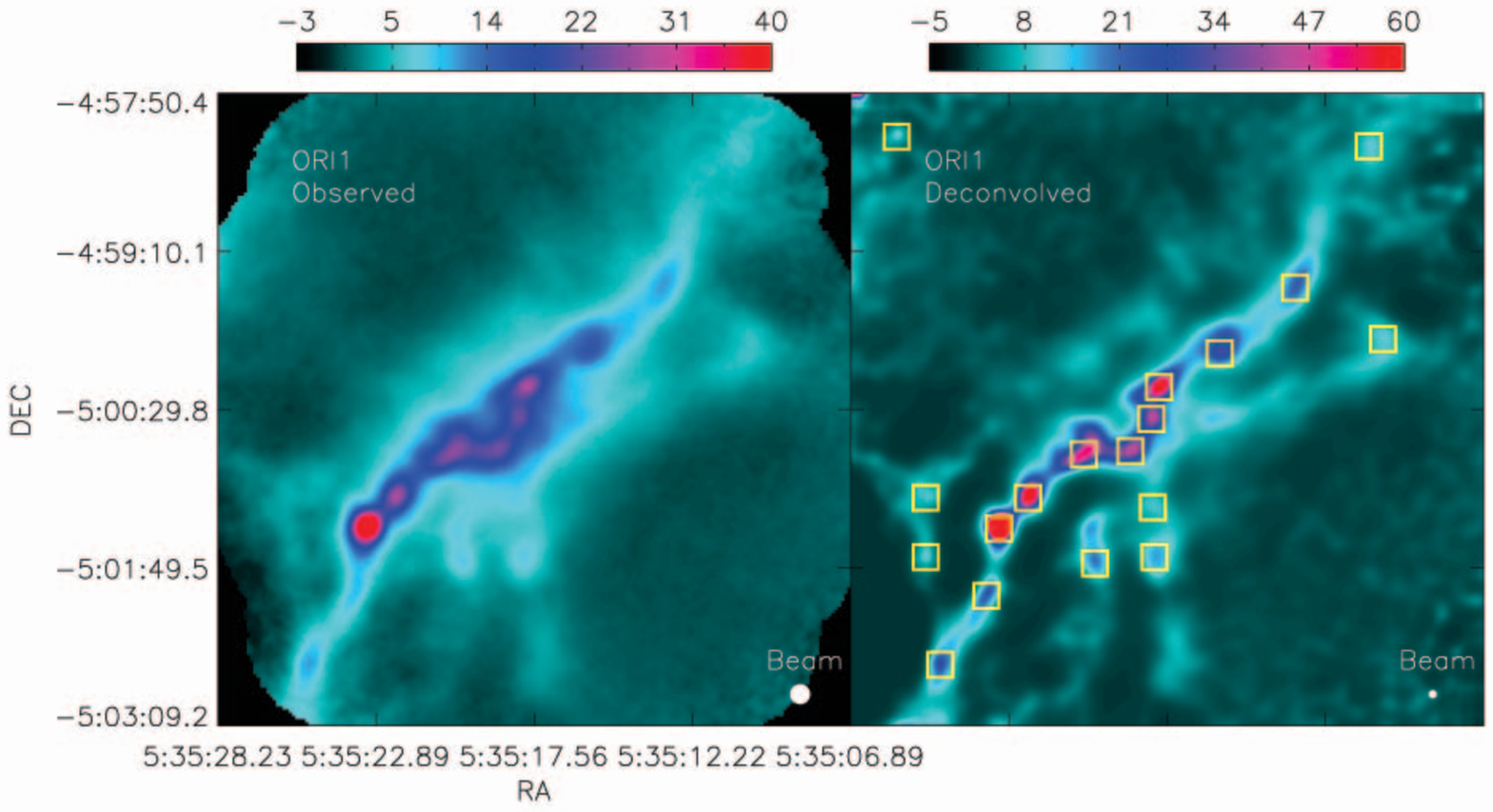}\\
   	\includegraphics[width=0.80\textwidth]{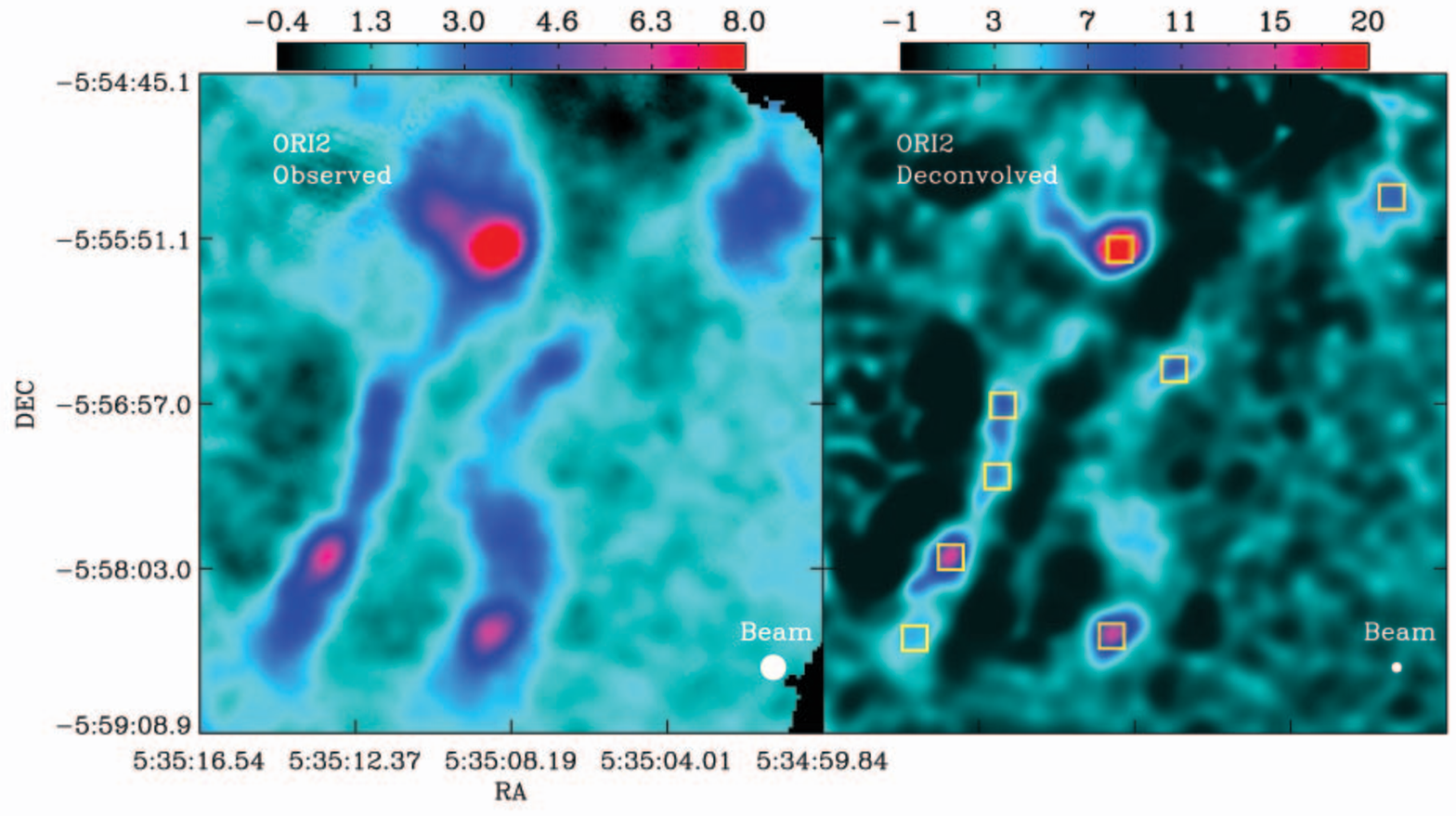}
   	\caption{Survey fields, ORI1, and ORI2.  Units and layout are the same as those for Figure 2. }
      	\label{fig:ori1-2}
\end{figure}

\begin{figure}[htp]
   	\includegraphics[width=0.80\textwidth]{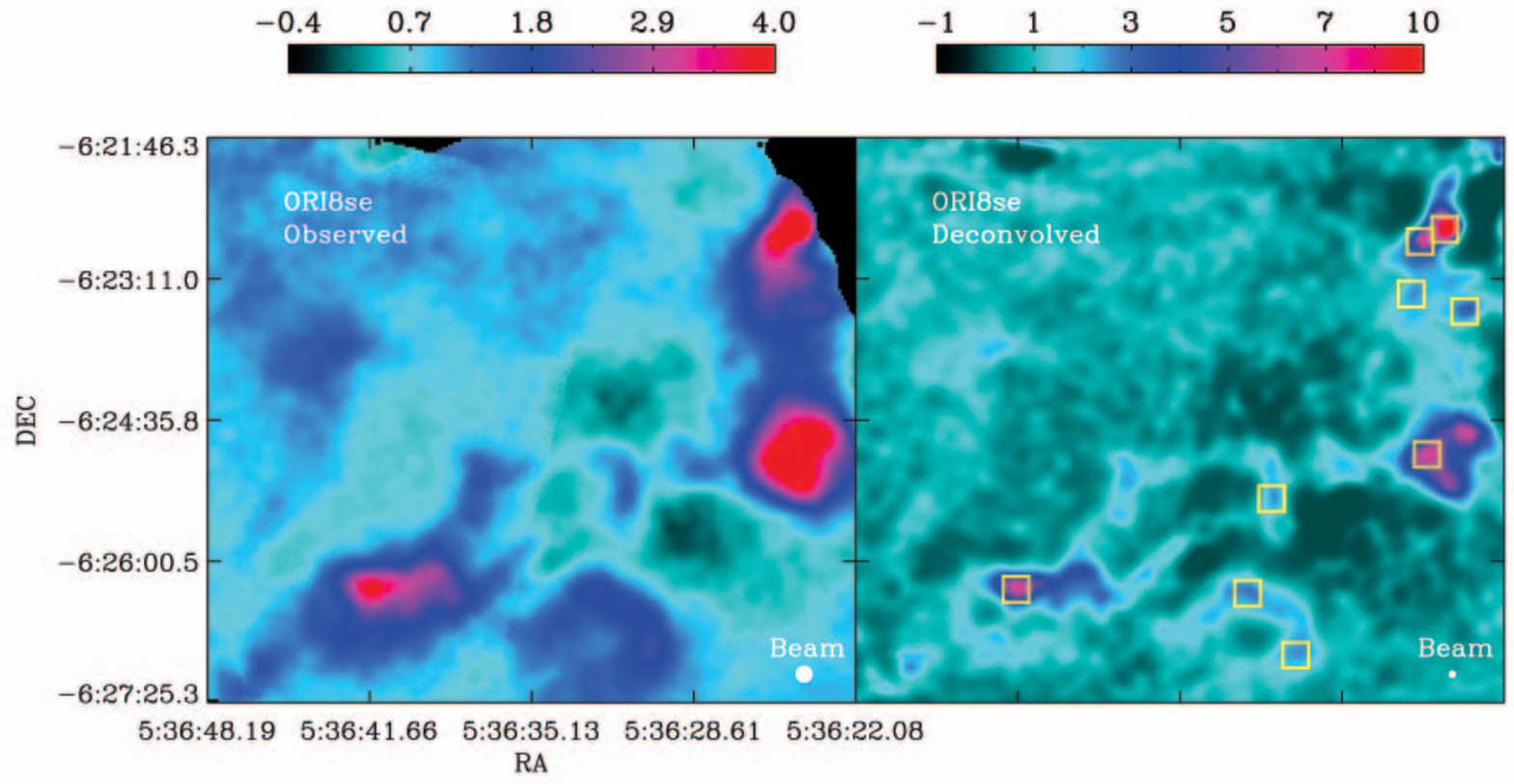}\\
   	\includegraphics[width=0.80\textwidth]{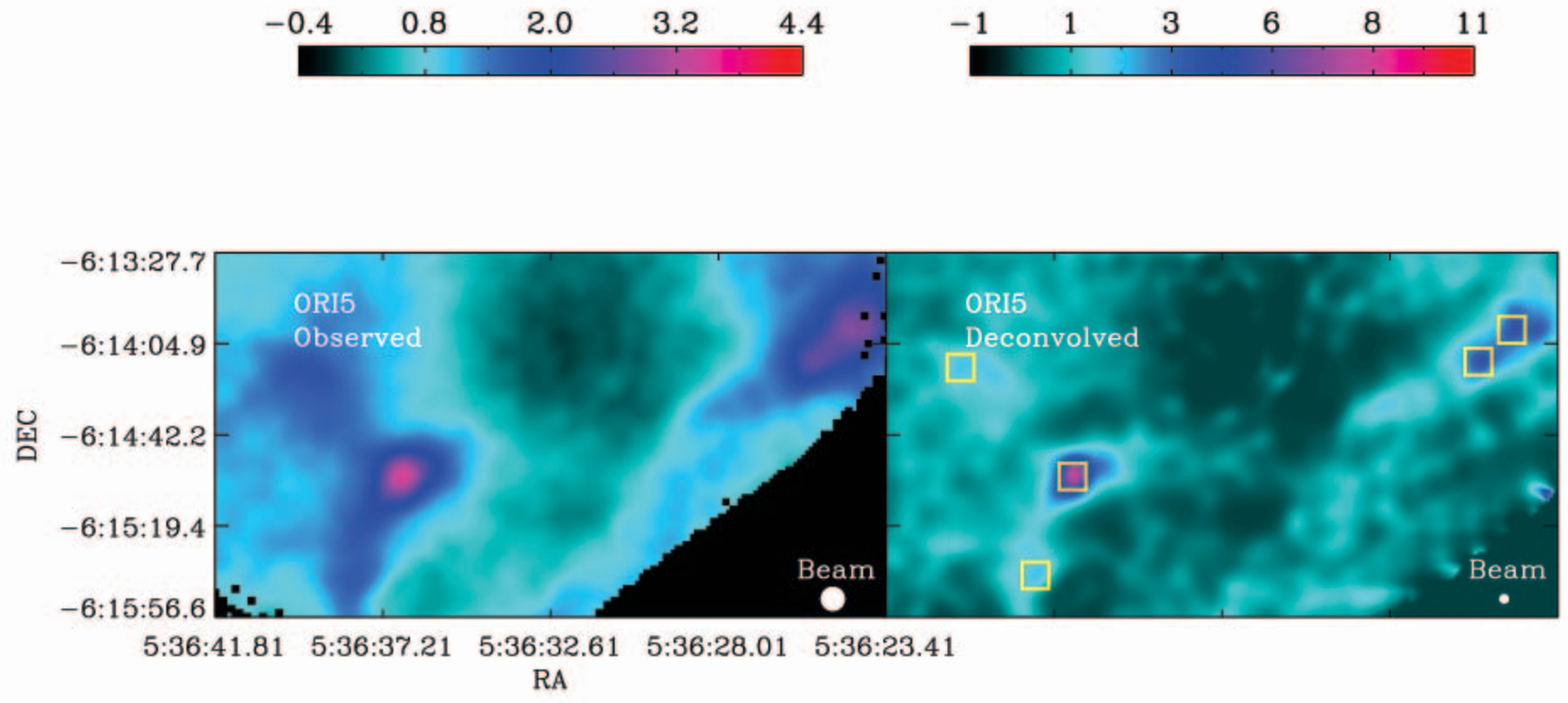}
   	\caption{Survey fields, ORI4, and ORI5. 
Units and layout are the same as those for Figure 2.}

      	\label{fig:ori45}
\end{figure}

\begin{figure}[htp]
   	\includegraphics[width=0.8\textwidth]{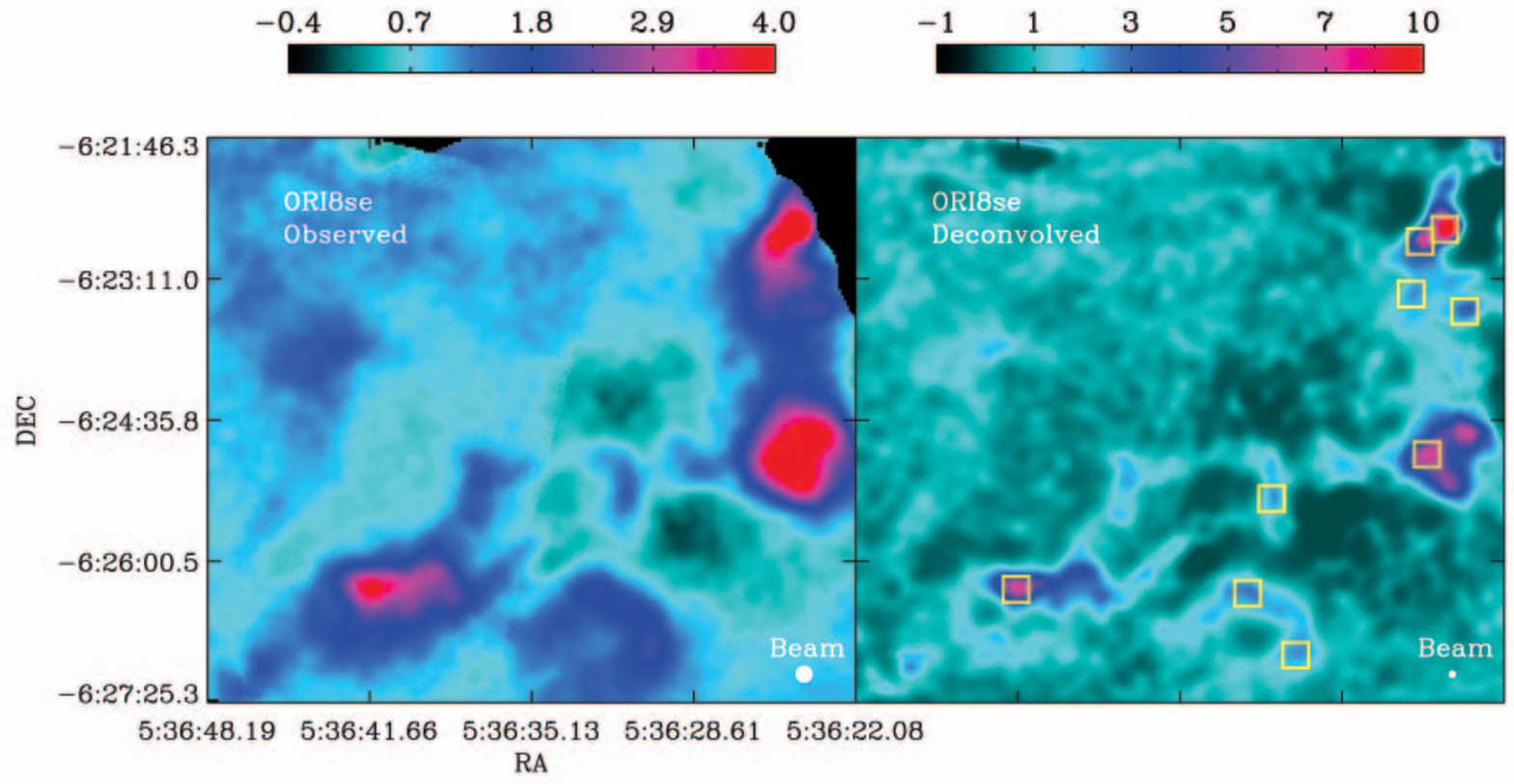}\\
   	\includegraphics[width=0.80\textwidth]{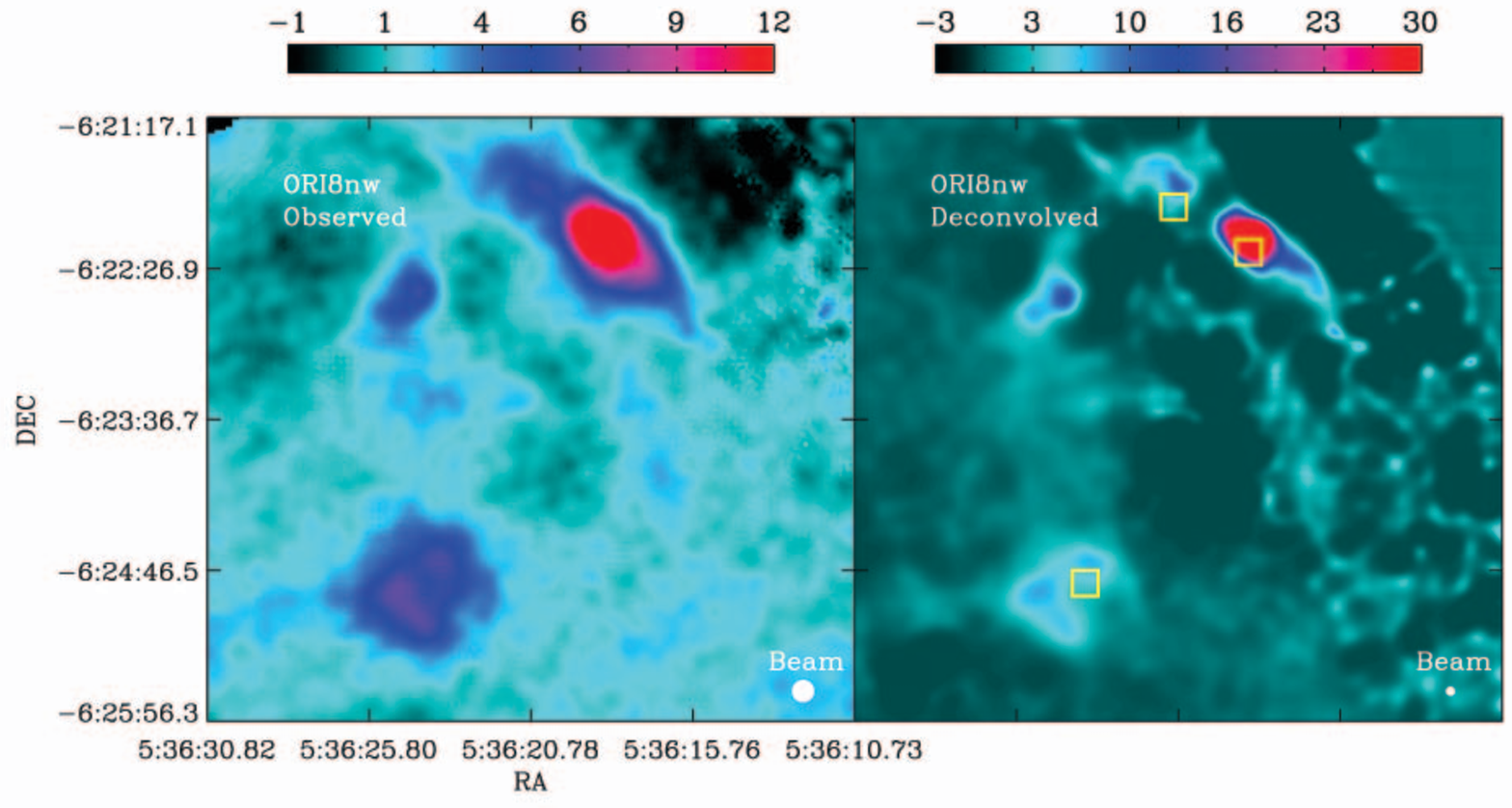}
   	\caption{Survey fields, ORI8SE, and ORI8NW. 
Units and layout are the same as those for Figure 2.}
      	\label{fig:ori8}
\end{figure}

The calibrated and deconvolved images are presented in Fig.~\ref{fig:ori7-11} through Fig.~\ref{fig:ori8}.
The Hires deconvolution conserves the total flux. 
When expressed in the same units of Jy per (9\arcsec)$^2$ beam as in the observed images, 
the peak values of the brightness of dust cores in the deconvolved image tend to be significantly higher than those in the original image.
This increase is expected because deconvolution sweeps the flux from the area covered by the extended error  beam pattern into the main beam.
It also helps recover condensations that are close to a strong source. 
Both these consequences of using Hires are important for better determination of the mass and the number count of cores. 
Another direct result of deconvolution is that the size of cores is generally smaller than in the original image. 
The best possible determination of the core size is important for evaluating the dynamical state of the cores, as will be discussed later. 

The detection of cores in our survey is limited both by the brightness of sources and their sizes.
If the source is a point source, then the requirement for a positive detection is a high signal to noise ratio, nominally  10$\sigma$, in the one resolution element that has signal. 
If the source occupies $N$ resolution elements, then a lower signal to noise ratio in each pixel is adequate.
To achieve the same statistical significance as that of the detection of the point source, the required signal to noise ratio in each pixel is scaled as $1/\sqrt{N}$. 
For an isothermal, optically thin dust core, the total continuum flux is linearly proportional to its dust mass.
Therefore, we can define a minimum detection mass $M_{det}$ for a core of a certain size to be 
\begin{equation}
  M_{det} = M_{point} \times \sqrt{N} \lc 
\label{det}
\end{equation}
where $M_{point}$ is the minimum mass of a point source (present only in one beam) to be detected. 
We note that $M_{det}$ scales linearly with the diameter of the core. 

In Fig.~\ref{fig:hist}, the minimum detection mass is plotted as a shaded area
in the histogram plotted against mass bins. 
The lower mass boundary at 0.04 \Ms\ of the region of incompleteness is set by a point source of 1.5 Jy, i.e., a 10 $\sigma$ detection of the peak pixel.
$M_{det}$ is calculated based on the same assumption of dust properties and cloud distance as used in the calculation of core masses (see \S\ 4) and a representative dust temperature of 17 K.
The upper mass boundary of the region of incompleteness, 0.6 \Ms, is set by a core of 0.35 pc diameter corresponding to about 2.5\arcmin\ angular size. 
This limit corresponds to the largest linear dimension of the SHARC II array. 
Judging by our maps, there exists no core larger than this size with structure so smooth that it is not resolved by SHARC II. 
Thus, the higher mass boundary represents a robust detection limit, above which the Orion core sample is complete.

\section{Characterizing the Cores}

We used a two step method to identify the cores in our data and obtain their physical parameters.
First, we employed Clumpfind (Williams et al. 1994) to obtain the intensity peaks. 
Clumpfind draws contours at specified levels and selects regions that are enclosed by these contours. 
In a two dimensional bolometer map, it is straightforward to assign a signal to noise ratio to the intensity peaks. 
We have used the following criteria to define an individual intensity peak --
it must be stronger than 10 $\sigma$ based on a noise estimate from empty regions and the valley between two peaks must be lower than one tenth of the stronger peak.
The intensity peaks thus obtained are also inspected to exclude isolated bright pixel spikes, which tend to appear toward edges
 of the maps, where the integration time is smaller than for the rest of the map. 
The end result is a list of clump candidates, for which the peak intensity pixels are at a significance level better than 10 $\sigma$, and which are well separated from nearby peaks.

To derive the clump parameters from these intensity peaks, we need to make some assumption about the boundary and shape of these structures. 
The assumption Clumpfind makes is to look for closed contour boundaries. 
In a region such as Orion where the core density is high and there is underlying diffuse structure, the intensity contours toward the edge of the clump are affected by emission not directly associated with individual clumps. 
The resulting core boundaries (contours) thus may have sharp corners and other strange shapes. However,
incorrectly defining the core boundaries usually has only a small effect on the derived total intensities, 
which are dominated by the central pixels as defined by our selection criteria.
But the sizes of cores can not be well defined this way.
Therefore, we make another assumption to help us define a core, which is that the core structure projected onto the sky is a two dimensional Gaussian.
Such an assumption enables us to fit the structure around an intensity peak by a well--defined Gaussian clump. 
We also simultaneously fit a flat background with the two dimensional Gaussian to account for the underlying diffuse structure. 
The majority of the cores are found to be nearly circular, thus
the clump size is defined by the average of the 1/e dimensions of the fitted Gaussians.
These fitted Gaussian prove to be good approximations to the two dimensional
brightness structures observed.
We present the characteristics of the 51 cores found in this study in Table 1.

\section{Calculation of Core Masses}

For an optically thin dust cloud, the dust emission can be described as (Hildebrand 1983)
\begin{equation}
S(\nu)=N(a/D^2)Q(\nu)B(\nu,T_d) \lc
\label{fdust}
\end{equation}
where $S(\nu)$, having units of erg~s$^{-1}$~cm$^{-2}$~Hz$^{-1}$ (or Jy), is the flux density produced by a cloud at distance $D$. 
It is the summation of the emission from $N$ spherical grains, with absorption coefficient $Q(\nu)$ and geometric cross section $a$. 
The formula is obviously true for point sources and is still applicable for extended sources as long as the dust seen in a telescope beam is emitting isotropically.

The value of the parameter $Q(\nu)$ and its dependence on frequency 
are affected by intrinsic properties of dust grains, such as their size
and composition. 
A simple power law ($Q(\nu)\propto \nu ^\beta$) model  has been predicted by theoretical work (Gezari, Joyce \& Simon 1973; Andriesse 1974) and has given reasonable fits to observations, but the spectral index $\beta$ varies from 0.6 to 2.8 (Wright 1987; Mathis \& Whiffen 1989; Lis \& Menten 1998). 
Moreover, $\beta$ itself may also depend on $\nu$. 
The general trend is that $\beta$ is smaller for higher frequencies 
(Hildebrand 1983; Draine \& Lee 1984; Martin \& Whittet 1990; Gordon 1995). 

In better studied GMCs, such as M17, $\beta \approx 2$ is usually a good description of observations (Goldsmith, Bergin \& Lis 1997). 
By combining continuum data at 350 \micron\ and 1100 \micron, Lis et al.\ (1998) find evidence that $\beta$ increases as the telescope beam moves away from
the Orion Bar, a photon dominated region (PDR), to more quiescent gas
further north. 
A larger $\beta$ in quiescent clouds is consistent with the hypothesis that  grains grow in size in such environments. 
Lis et al.\ indicate that dust temperature $T_d = 17$ K and $\beta = 2.5$
for regions near ORI1. 
Their maps do not cover other regions in our survey, which are located south of the Orion Bar. 

With the spectral index known, the actual value of $Q$ can be estimated and compared with `standard' values
 measured at other wavelengths. 
Extrapolating the 125 \micron\ emissivity of Hildebrand (1983) with $\beta=2$ gives $Q(350)=1\times10^{-4}$. 
Extrapolating the 1300 \micron\ value of Chini et al.\ (1997) gives 
$Q(350)=2.2\times10^{-4}$. 
Direct measurements of part of the Orion molecular cloud by Goldsmith et al.~(1997), with an assumed  gas to dust ratio equal to 100 and dust temperature 
$T_d=17$ K, give $Q(350)=4\times10^{-4}$. 
We use a representative value of $Q(350)=2\times10^{-4}$  in this paper.
The range of dust emissivity as discussed above is indicative of the uncertainties in determining the dust mass from dust continuum emission, due to the complexity involved in modeling $Q(\nu)$.

The temperature of dust grains is another issue in deriving the dust mass. 
In star forming regions, the dust temperature is determined by energy equilibrium between UV/optical absorption and infrared emission. 
The fact that grains can be of different sizes dictates a distribution in grain temperature. 
The existence of large, cold grains makes the dust mass derived 
from dust emission based on single temperature fits an underestimate (Li, Goldsmith, \& Xie 1999). 
In well--shielded regions, the dust temperature will also be affected by gas--dust coupling.  
For lower densities (\nh$<10^{5}$ \cc), the dust temperature will be a few degrees lower than the gas temperature, according to the modeling by Goldsmith (2001). 
At \nh=$10^{6}$ \cc, $T_d \approx T_{gas}$. In our cores, the 
average gas density estimated based on data presented in Table 1 is generally
greater than $10^{6}$ \cc. It is thus reasonable to use the gas temperature
as a direct estimate of dust temperature.

We have good measurements of the gas temperature from \ammonia\ observations for most of the survey regions, which are at $\sim$ 1\arcmin\ resolution (Paper I).
The one $\sigma$ statistical uncertainty of our temperature measurements due to data noise is 0.9 K. 
The temperature in regions not covered in our ammonia survey (ORI7 and ORI11) can be estimated based on their distances to the Trapezium cluster. 
Because these regions are already more than 8 pc away from the main ionizing source, the uncertainty in the temperature estimate due to the external UV field (the main heating source) is smaller than $\sim$ 1.5 K (Paper I and Stacey et al.~1993). 
Without strong embedded sources and external heating from outside (as is the case for most of our surveyed Orion region), the temperature will drop toward the cloud centers. 
But the temperature decrease is small for most of the cores, as the external UV field for our selected regions is relatively weak compared to that in regions closer to the center of Orion.
We will thus take the dust temperature to be equal to the gas temperature in the following discussion. 
This approximation is accurate for dense regions and an overestimate of $T_d$ for others.
Due to this possible higher than true $T_d$, the dust mass we determine may
be underestimated.

\begin{figure}[htp]
   \plotone{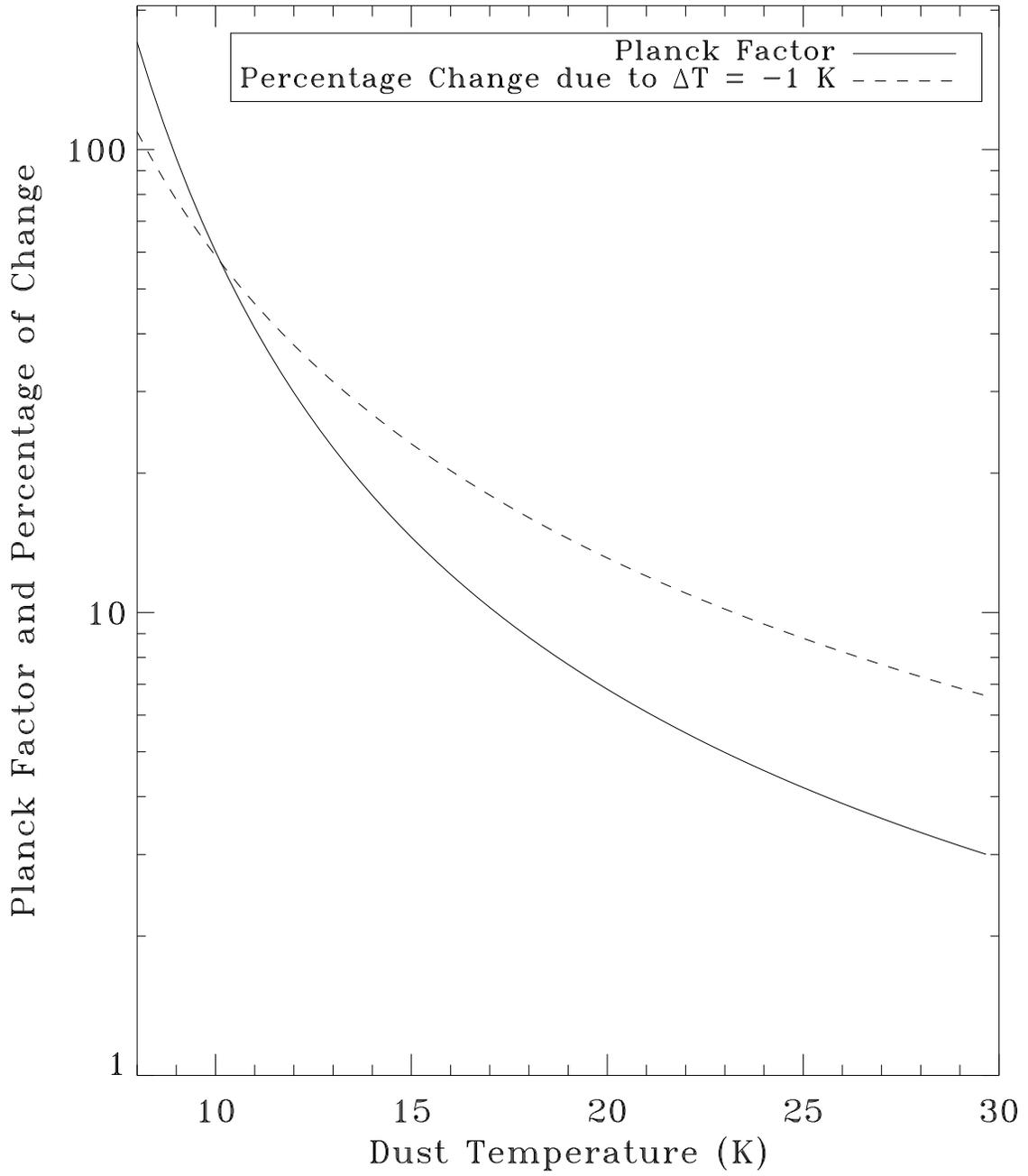}
    \caption{The Planck factor and its percentage change, calculated at a wavelength of 350 \micron\, due to a 1 K decrease in the dust temperature.}
      \label{fig:td}
\end{figure}

Assuming a grain radius $r=0.1$ \micron, a grain density $\rho = 3$ g \cc, cloud 
distance $D=480$ pc, gas to dust ratio $GDR=100$, and $Q(350)=2\times10^{-4}$
we can rewrite Eq.~\ref{fdust} in units more convenient for this situation
\begin{equation}
M_{core}  = 2.4 \times 10^{-2} M_\odot \Bigl[\frac{2\times 10^{-4}}{Q(350)}\Bigr]
\> \Bigl[\frac{\lambda}{350 \mu m}\Bigr]^3 \> \Bigl[\frac{D}{480pc}\Bigr]^2\>\Bigl[\frac{GDR}{100}\Bigr] 
\> \Bigl[\frac{S(\nu)}{Jy}\Bigr] \> P_f (T_d) \lp
\label{nd}
\end{equation}
 The Planck factor, $P_f(T_d)=e^{h\nu/kT_d}-1$, is plotted in  Fig.~\ref{fig:td}. 
We also plot the percentage change of the Planck factor if the dust temperature were to decrease by 1 K.

At a temperature of 15 K, a decrease of 1 K in the dust temperature corresponds to a 23\% increase in the Planck factor. 
The fractional change drops to 13\% at 20 K.
For ORI1, even if the gas and dust temperatures are not closely coupled, the uncertainty produced by using the gas temperature in deriving the dust mass should not be large thanks to the relatively high temperatures. 
For colder sources, the knowledge of the dust temperature becomes crucial since the Planck factor diverges toward lower $T_d$. 
The lowest temperature used in our calculation is 12 K, at which the 
uncertainty in temperature corresponds to about 40\% uncertainty in the derived core mass.
Other than the dust emissivity, this factor is the largest source of uncertainty in
our calculation.
The mass of each core is given in Column 6 of Table 1.

The mean core mass of our sample, 9.8 \Ms, is about 10 times larger than those found in low mass star forming regions and
isolated dark clouds (e.g.\ Benson \& Myers 1989; Young et al.\ 2006). 
We believe this difference is not due to the greater distance of Orion than the well studied LMSF regions. 
The Benson \& Myers (1989) study include both isolated cores and a collection of cores in Taurus. 
The median FWHM core diameter in that sample is 0.14 pc, which is about seven times the spatial resolution of SHARC II used in the present work ($\sim0.02$ pc at the distance of Orion). 
The BOLOCAM instrument used in Young et al.\ (2006) has a beam size of about 28\arcsec, which corresponds to about 0.019 pc at a distance of 130 pc.  
This is about the same as the spatial resolution of SHARC II for Orion.  
The average separation of sources in Young et al.\ (2006) is about 0.09 pc, significantly larger than our resolution. 
In short, the higher mass of the cores in our sample is unlikely to be a consequence of
their being combinations of multiple low mass cores such as those characterizing the Ophiuchus and Taurus regions.

\section{Core Stability}

Without spectroscopic data at a spatial resolution that matches that of the submillimeter continuum, we cannot determine 
the balance between gravity and turbulence and/or thermal support.
We can, however, examine the limiting case of the cores being only thermally supported.

For a spherical, self-gravitating, isothermal, and hydrostatically supported core, its density profile can be described by a family of solutions to the Lane-Emden equation (often called Bonnor-Ebert spheres: Ebert 1955 and Bonnor 1956), with the dimensionless radius being
\begin{equation}
\xi = r \sqrt{4 \pi G \rho_c/v_s^2}  \lc
\label{xi_def}
\end{equation}
where  $\rho_c$ is the central density and $v_s = \sqrt{kT/(\mu m_H)}$ is the sound speed.
Each solution is characterized by a single parameter $\xi_{max}$, 
which is determined by the outer radius 
and central density. This represents a truncation of the infinite isothermal sphere and 
the density profile within is thus fixed by the core size and central density (see discussion by Alves, Lada \& Lada 2001).

Tafalla et al.~(2004) have studied an analytic approximation to the density profile of Bonnor Ebert spheres
\begin{equation}
\rho(\xi) = \frac{1}{1+(\xi/2.25)^{2.5}} \lc
\label{xi}
\end{equation}
and have shown that it is within 10\% of the numerical solution 
given by Chandrasekhar \& Wares (1949) for $\xi < 23$.
Ebert (1955) and Bonnor (1956) pointed out that when $\xi$ exceeds a critical value of 6.5, the solution becomes unstable. 
Assuming $\xi_{max} = 6.5$ and using the observed core radius, a critical core mass $M_{BE-CR}$ can be calculated
by integrating Eq.~\ref{xi}.

\begin{figure}[htp]
   \plotone{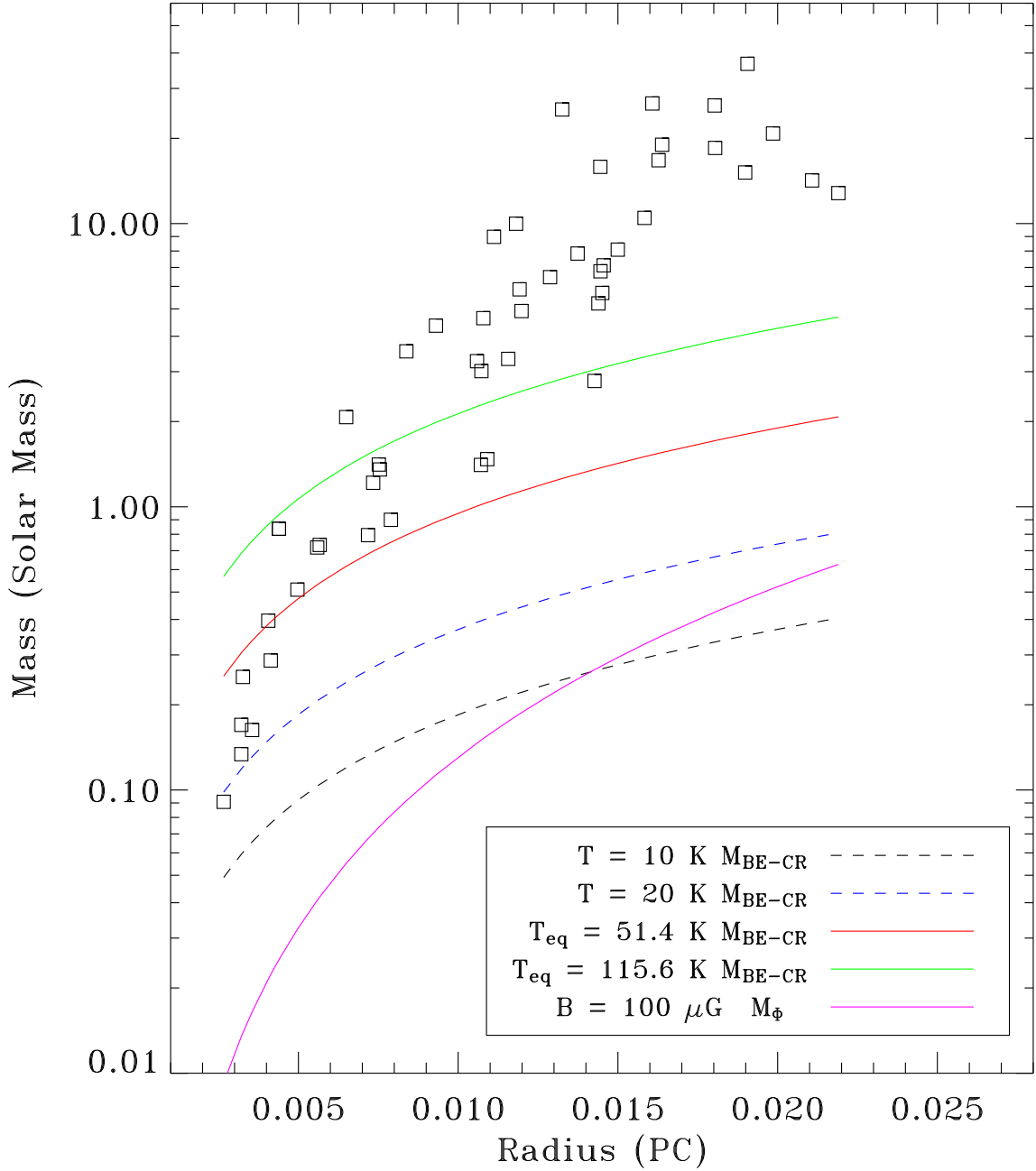}
    \caption{The mass and radii of the cores in our sample. 
The curves are the critical mass that can be supported for a core of specified temperatures and size.
$T_{eq}$ = 51.4 K corresponds to a turbulent line width of 1 \kms (FWHM), and 115.6 K corresponds to a line width of 1.5 \kms.  
$M_{BE-CR}$ is calculated based on Eq.~\ref{xi}. $T_{eq}$ is defined in Eq.~\ref{teq}. $M_\Phi$ is defined in 
Eq.~\ref{mphi}. }
      \label{fig:m2r}
\end{figure}

In Fig.~\ref{fig:m2r}, the sizes and mass of the cores are plotted along with the calculated curves based on critical Bonnor-Ebert spheres at various temperatures.  
The majority of our cores appear to be too massive to be stable Bonnor-Ebert spheres for a kinetic temperature as high as 30 K. 
The caveat for such a discussion is that the observed size of our cores are not
as clearly defined as required by the theory. 
But the large excess of the observed core mass relative to the Bonnor Ebert critical mass suggest that increasing the core size by as much as a factor of a few would not make a core thermally stable.

To evaluate the importance of turbulence, we can also consider a modified BE sphere taking into
account turbulence (cf.\ Lai et al.\ 2003).
We can define an equivalent temperature
\begin{equation}
T_{eq}=\frac{m_H \Delta V^2 }{8\ln(2)  k} \lc
\label{teq}
\end{equation}
where $\Delta V$ is the total full width half maximum (FWHM) line width of the gas.
The observed FWHM line width of \ammonia\ at 1\arcmin\ scale toward these regions ranges from 
0.7 km/s to 1.4 km/s (paper I). We have plotted the
critical mass curves based on $T_{eq} (\Delta V = 1$ km/s) and $T_{eq} (\Delta V = 1.5$ km/s) in Fig.~\ref{fig:m2r}. 
A turbulent line width of 1.5 km/s or more would in principle offer  significant support.
However, given  the generally quiescent state of such cores and the decrease of turbulence from low density to high density regions (Paper I), one would not expect the turbulence within our cores to exceed the value observed at the larger (1\arcmin) scale.

We do not have spectroscopic data with spatial resolution matching that of the present submm continuum data. 
The relevant extant data is from ammonia (Paper I), giving the linewidth
at $\sim$50\arcsec\ scale, which is substantially larger than the core sizes. 
The ammonia linewidth (which is generally two to three times larger than the linewidth found in the cores 
studied by Benson \& Myers 1989) seems to indicate approximate virial equilibrium between 
the random turbulent motions in these cores and their self--gravity.
If the turbulence decreases at smaller scales as suggested in paper I, 
then it would not be sufficient to stabilize the cores, and they would be candidates for collapse and the onset of star formation. 

Finally, we evaluate the importance of magnetic pressure.
For a cloud with uniform density and uniform magnetic field, 
the maximum mass which can be supported by a steady $B$ field alone can be derived
through virial theorem (e.g.~Spitzer 1978)
\begin{equation}
M_\Phi=\Bigl[\frac{5}{9G}\Bigr]^{1/2} Br^2  \lc
\label{mphi}
\end{equation}
i.e., the magnetic supported mass is proportional to the flux.
It is easy to see that if the cloud is conducting thus freezing the magnetic flux, 
the contraction of a cloud cannot proceed if the starting cloud mass is smaller than $M_\Phi$.
Such a uniform cloud combined with isothermal conditions cannot be 
in pressure equilibrium. Numerical simulations have been carried out to 
study magnetized clouds under more realistic conditions. The evolution of such a system
is complicated, depending on the initial mass, geometry, and the threading
of magnetic field. For our purpose of estimating the overall importance of a steady field,
we note that simulations for spherical magnetized clouds (Mouschovias \& Spitzer 1976; 
Tomisaka, Ikeuchi \& Nakamura 1988) give the critical mass of a cloud in equilibrium 
to be of the same form as in Eq.~\ref{mphi}, modified by a correction factor 
$M^\prime_\Phi = c_\Phi M_\Phi$. 
The correction factor, $c_\Phi$, is shown to be 
smaller than unity, as the centrally enhanced density profile of
a cloud in pressure equilibrium makes the cloud easier to 'squeeze' from outside.
We thus use Eq.~\ref{mphi} to estimate the maximum mass that can be supported by a steady
magnetic field.

The measurement of the magnetic field strength is difficult.
There exists one such measurement in the Orion molecular cloud.
Using the IRAM 30m telescope, Crutcher et al.\ (1999) detected the Zeeman effect in the CN 3mm line near Orion BN/KL. 
The field strength is derived to be either 190 $\mu$G or 360 $\mu$G depending on the fitting scheme. 
This is much larger than the $B\sim30 \mu$ G measured in dark clouds (e.g.~Goodman et al.\ 1989). 
Given the proximity of BN/KL to active star formation, the large value of $B$ could be explained by a rapid collapse freezing the magnetic flux into high density regions along the line of sight.  
It is not clear how different the magnetic field should be in our quiescent cores, but it is expected to be smaller than the Orion KL values measured by Crutcher et al.
We thus take a nominal $B=100\>\mu$G in our calculations of $M_\Phi$.
The resulting $M_\Phi$ is also plotted in Fig.~\ref{fig:m2r}. 
A static B field of this magnitude adds only minimal support and even the higher value of the magnetic field measured by Crutcher et al.\ (1999) would not be significant.

To determine the dynamical state of these cores more accurately will require higher angular resolution spectroscopic data for the gas as well as better determination of the magnetic field strength and morphology.
At this point we can say that the large mass of the Orion cores derived in section 4 
suggests that these cores are either collapsing or are supported by strong turbulence.

\section{Core Mass Function}

\begin{figure}[htp]
  \plottwo{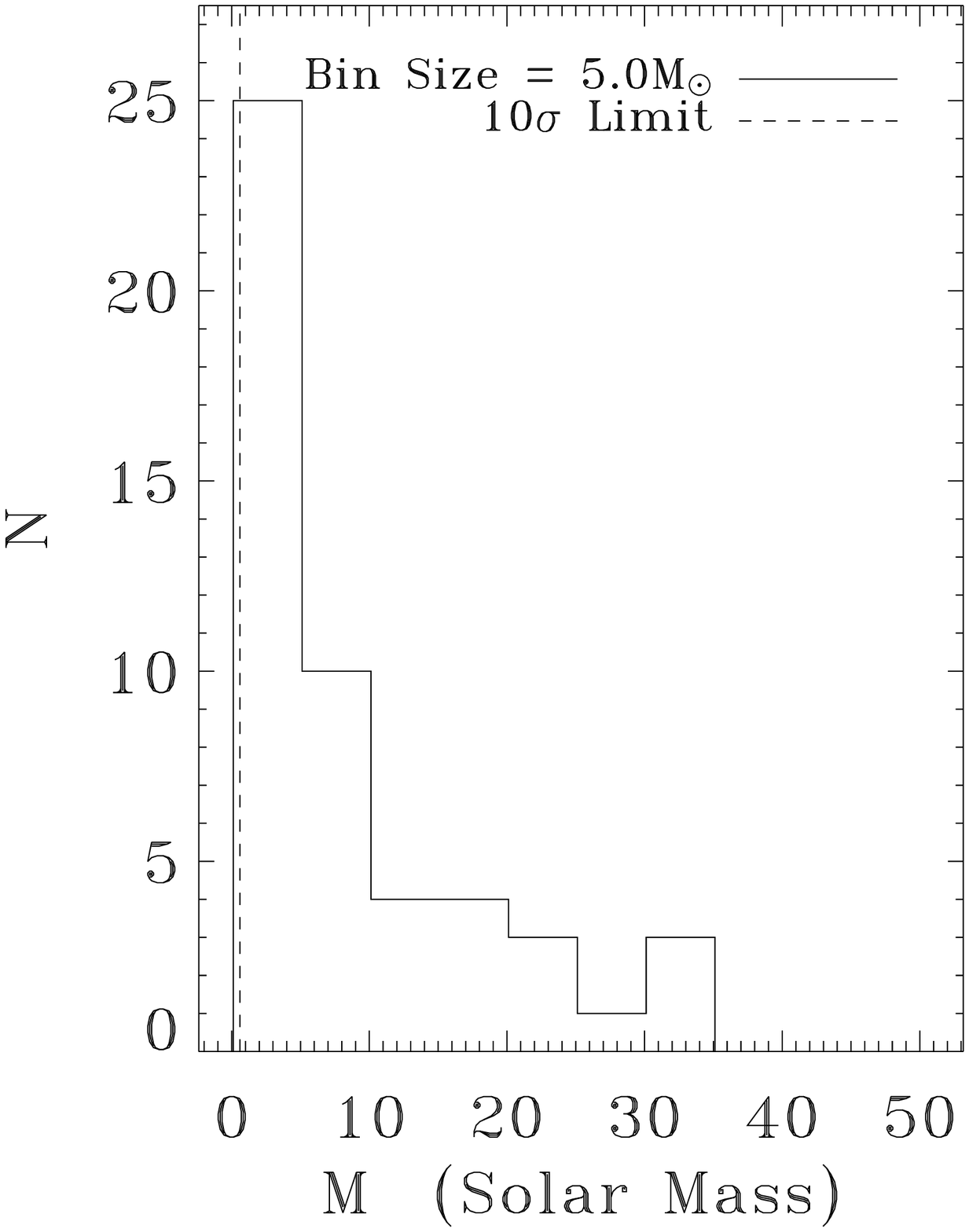}{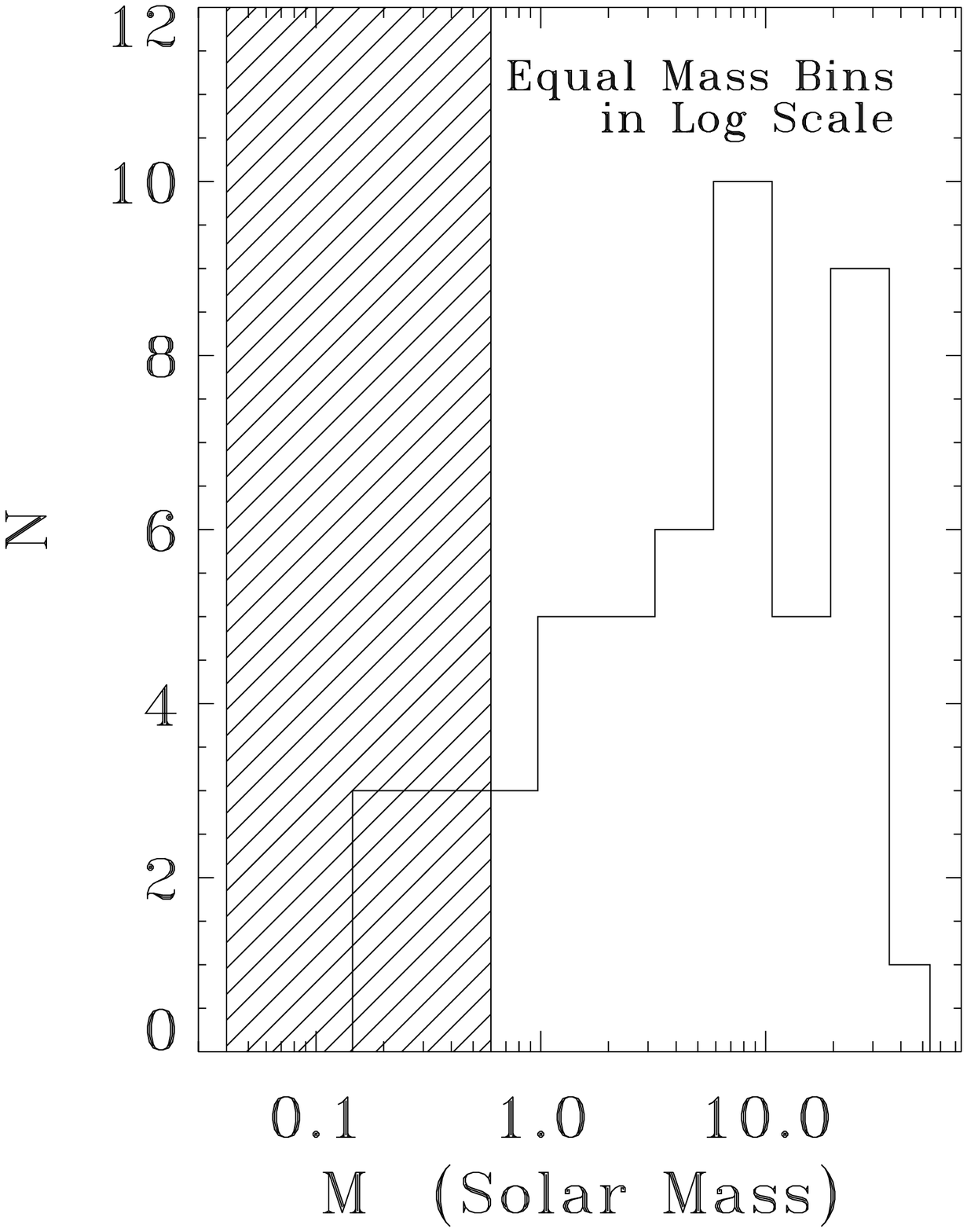}
  \caption{Histogram of core masses in our sample. 
Left: Linear mass bins. Right: Logarithmic mass bins.
The shaded area represents the region of incomplete sampling as discussed in the text.}
      \label{fig:hist}
\end{figure}

Based on the core masses derived, we present the statistics of the core masses in Fig.~\ref{fig:hist}, left panel.
This figure displays the differential mass distribution, in which the data are binned into mass ranges of specified size.
When grouped in mass bins of uniform width 5 \Ms, the number of cores within each mass bin drops from 26 in the lowest mass bin (0 to 5 \Ms) to 1 in the highest mass bin (40 to 45 \Ms). 
The incompleteness of the survey is in the lowest mass bin. 
To get a better look at the distribution of core masses at lower masses, the histogram is also plotted in terms of mass bins having equal logarithmic widths (Fig.~\ref{fig:hist}, right panel).

We would like to be able to represent the distribution of core masses by some simple function, which is generally referred to as the core mass function, or CMF.  
In the following we show that determining a robust CMF from the core mass distribution is not a trivial undertaking.
There are two widely used approaches to fitting an analytic function to the core mass distribution in order to determine the CMF.  
These are to employ (1) the cumulative mass function, which considers the fraction of cores having mass greater than some specified value, and (2) the differential mass function discussed above. In this paper, CMF refers to the differential
core mass function as defined in Eq.~\ref{dndm} in the Appendix unless specified otherwise.

\begin{figure}[htp]
\plottwo{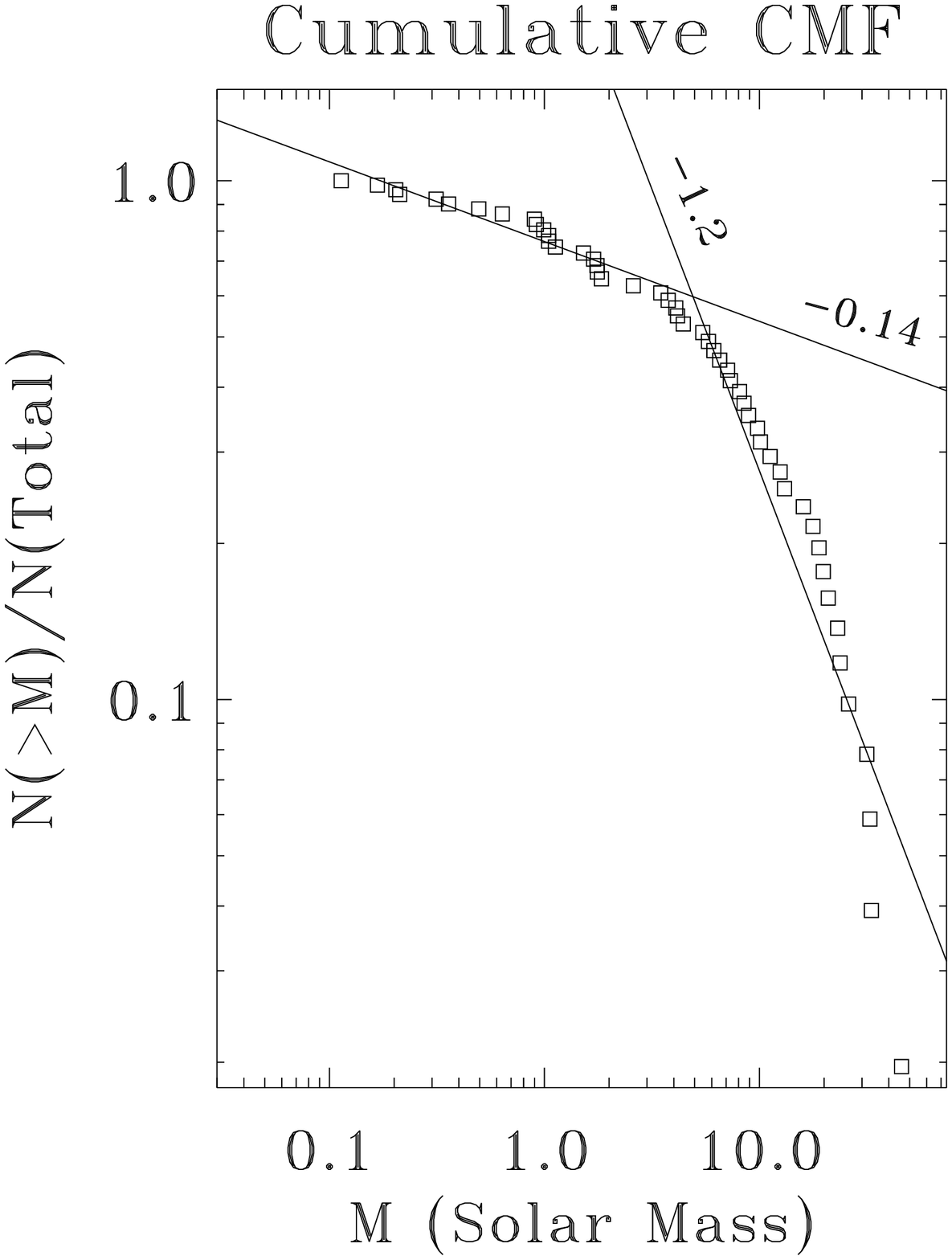}{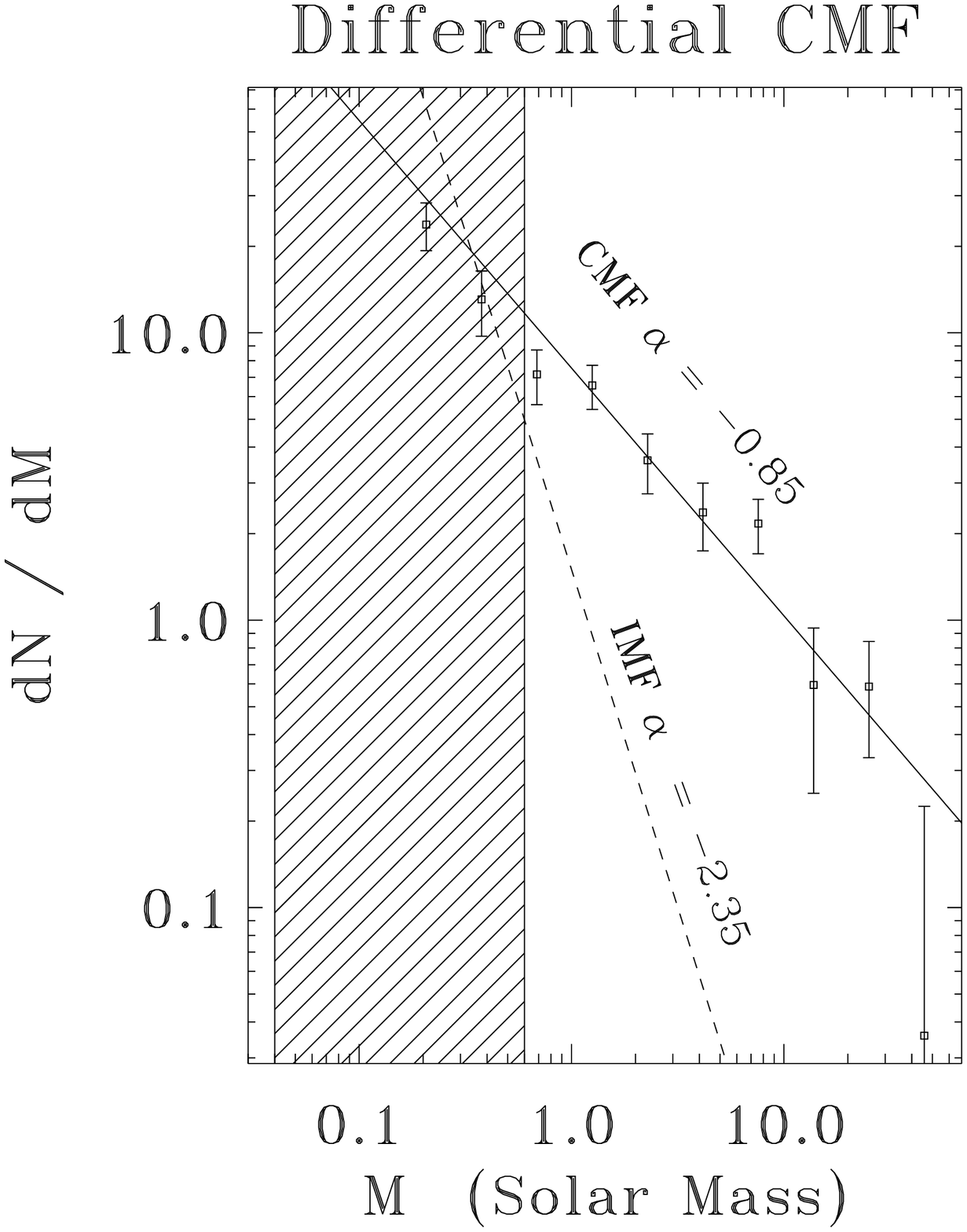}
    \caption{Left: Cumulative Core Mass Function with the two power laws that are fitted to the different portions of the data. 
The power law index for each fitted line is also labeled in the plot. 
Right: Differential Core Mass function.
The dashed line shows the slope of a Salpeter stellar IMF and the solid line is the best fit to the binned differential CMF with a slope equal to -0.85. 
The shaded zone indicates the region where the core sample may be incomplete; it is the same as in Fig.~\ref{fig:hist}.}
      \label{fig:cmf}
\end{figure}

We present both types of mass functions in Fig.~\ref{fig:cmf}.
The cumulative CMF appears to suggest two power laws:  a steep power law with an equivalent index of -2.2 for the larger mass cores and a flatter power law of slope -1.1 for the lower mass ones.
The fitting results based on the assumption of two power laws are similar to results from several past studies (e.g.\ Reid \& Wilson 2005 and references therein).
However, as we will discuss in detail in the Appendix, fitting multiple power laws to the cumulative mass function is
prone to ambiguous interpretation, especially when the power law index of the CMF is about -1. 

For the Orion cores, the single power law fits based on the differential CMF yield indices close to -1. 
One example based on 10 bins is plotted in the right hand panel of Fig.~\ref{fig:cmf}, 
which has a best fit index of $\alpha$ = -0.85$\pm0.21$.
If we change the number of bins, the exact value of fitted index varies between about -0.8 to -0.95.
These values are close to -1 and any two power law  interpretation of the cumulative CMF must therefore be
considered with some caution.

To understand better the consequences of different fits to the cumulative CMF, we have constructed simulated core samples and analyzed them to evaluate different fitting procedures.
The cores are randomly generated within the same range of mass as that of the observed Orion core sample and the probability density function according to which a core will have certain mass is based on a power law.
In Fig.~\ref{fig:sim}, the cumulative CMF of the Orion cores and a family of simulated core samples are shown together with our CMF determined for the Orion cores. 
Each simulated sample has 1000 cores and the true mass functions for the samples are power laws with indices ranging from 0 to -2.0. 
Fig.~\ref{fig:sim} makes it obvious that the cumulative CMF plotted in log--log space has curvature for power law indices larger (flatter) than -1.5. 
In the extreme and illustrative case of $\alpha = 0$, the underlying CMF is strictly flat, i.e., the same probability of detecting high mass cores as low mass ones. 
The cumulative CMF, however, has a clear turnover, as any sampling 
(or observation) of a probability density function is done in a limited mass range. 
This should already sound an alarm for any direct power law fit to a cumulative function as a way of modeling the underlying CMF.

When sampled somewhat sparsely, the cumulative CMF can easily be mistaken to be characterized by two regions each having a different power law index, with the low mass end being flatter and the high mass end being steeper. 
However, an important result here is that the simulated CMF which best agrees with that of the observed cores has a power law index of -0.80. 
This value is consistent with the result from the fit to the differential CMF given above.
The advantage of using the cumulative mass function lies in the fact that the number of data points is equal to the number sources in the sample.  
That number will obviously be greatly reduced when binning is required, as is the case for the differential function. 
In the case of the index being close to -1, however, the differential mass 
function is a much more reliable method than the cumulative mass function to derive the CMF. 

In fact, when comparing with other massive regions, such as Orion B (Johnstone et al. 2001), NGC 7538 (Reid \& Wilson 2005) and RCW 106 (Mookerjea et al.~2004), a power law index close to -1 can be derived from their core samples when only those  cores in a mass range similar to ours are included. 
The steeper power law indices ($\sim$ -2.3) based on those samples are usually a fit to the higher mass portion of their cores (see the review in Reid \& Wilson 2005).
Due to poorer angular resolution and sometimes larger distances, the high end of core mass in these samples can be as large as tens of thousands of \Ms\ and
some contain water masers or bright infrared sources, which are signs of active star formation. 
One should use caution when interpreting these dense structures together with resolved cores, especially when multiple power laws are fitted to a cumulative CMF.

From a theoretical viewpoint, Padoan \& Nordlund (2002) give an analytic relation between the CMF index and 
the power spectrum index, %\beta$,  of the supersonic turbulence. 
Such a relation would predict a stellar IMF--like CMF for core masses larger 
than 1 \Ms\ with $\beta = 1.74$, consistent with some observations (Miesch \& Bally 1994) , but different
from others (Brunt \& Heyer 2002).
For lower mass cores, these authors state that the mass distribution 
will be flatter as the number of gravitationally unstable cores drops.
Simulations (e.g.~Gammie et al.~2003, Klessen \& Burkert 2001, Tilley \& Pudritz 2004) under varying conditions can produce core samples with a CMF consistent with the Salpeter IMF, although this conclusion is not very restrictive and may depend on the evolutionary time of the observation (see discussion by Gammie et al.~2003).
A very recent numerical study (Ballesteros-Paredes et al.~2006) finds that the 
CMF is dependent on the sonic Mach number and that the CMF in a supersonic turbulent flow may have changing slopes as a function of time. 
Although different in their opinions regarding the direct relationship between 
IMF and CMF, these studies all relate the CMF to the turbulence conditions in the clouds included in their calculation.
In order to compare theoretical predictions of the CMF with observations, it is important to have similar definition and scales of structure while assembling the statistics. 
The Orion cores are likely to be supercritical (section 5), are quiescent, and are probably precursors of protostars. 
The flat CMF of Orion cores suggests that evolution and shaping of the mass distribution of dense gas continues after collapse has already started.

Based on consideration of both the cumulative and the differential CMF, 
we find that the CMF of the Orion cores has a significantly flatter CMF than that of low mass star forming regions.
This may reflect the different environment in these regions. 
In particular, as the Orion CMF is also significantly flatter than the stellar IMF, one would expect environmental effects in the later stages of star formation to shape the IMF, as it cannot be a result of collapsing each core with a similar star formation efficiency.

\section{Conclusion}
We have identified 51 dust cores in a 350 \micron\ submillimeter continuum survey of the quiescent regions of the Orion molecular cloud using the SHARC II camera. 
The enhanced spatial resolution of our data using the Hires deconvolution tool and our knowledge of the temperature from ammonia mapping (Paper I) enable us to determine relatively accurately the number, size and the total mass of these cores. 
This Orion dust core sample: 
\begin{enumerate}
\item is a collection of resolved or nearly resolved cores, with a mean mass of 9.8 \Ms\, which is one order of 
magnitude higher than that of resolved cores in low mass star forming regions;
\item includes largely thermally unstable cores, which are unlikely to be stabilized by the magnetic field,
suggesting that the cores are supported by strong turbulence or are collapsing; 
\item has a power law core mass function with index $\alpha =$ -0.85$\pm 0.21$, 
which is significantly flatter than the stellar IMF and than that found for core samples in low mass star forming regions.
\end{enumerate}

Our comparison of the use of differential and cumulative mass functions to analyze the core mass distribution indicates that the differential approach, while requiring more cores due to the binning involved, is more robust and has better defined statistical uncertainties. 
The cumulative mass function approach can erroneously suggest multiple power law indices, particularly if the underlying core mass distribution is characterized by a power law index $\simeq$ -1.

\appendix
\section{Appendix: Differential and Cumulative Function for Power Law Distribution}

Following the convention used for the stellar initial mass function, we can define a power law core mass function (CMF) for cores as 
\begin{equation}
\frac{dN}{dM} \sim M^{\alpha} \lc
\label{dndm}
\end{equation}
where the number of cores $dN$ in certain mass range $dM$ is a power law of index $\alpha$.
The cumulative CMF is then
\begin{eqnarray}
\nonumber N(>M_0) &=& \int_{M_0}^{M_{max}} dN \\
     &\sim& \frac{1}{\alpha+1}M_{max}^{\alpha+1}-\frac{1}{\alpha+1}M_{0}^{\alpha+1} \lp
\label{nm}
\end{eqnarray}
where $M_{max}$ is the maximum mass for a certain sample. 

It is obvious that in an ideal case, the cumulative CMF will also be a power law, but with a flatter slope. 
In the literature, the cumulative CMF plotted on a log--log scale has been fitted directly by straight lines to give one or multiple power law indices. 
The indices are then decreased by one to give the CMF power law indices as defined in Eq.~\ref{dndm}. 

Such direct fit does not work when the power law index is close to -1. 
When $\alpha =$ -1, the integral in Eq.~\ref{nm} results in an log function, $log (M_{max}) - log (M_0)$ , {\em not}  a power law.
Instead of a single slope on a log--log plot, the appearance of the cumulative mass function has
curvature, even when the underlying core mass distribution can be
characterized by of a single power law.

\begin{figure}[htp]
   \plotone{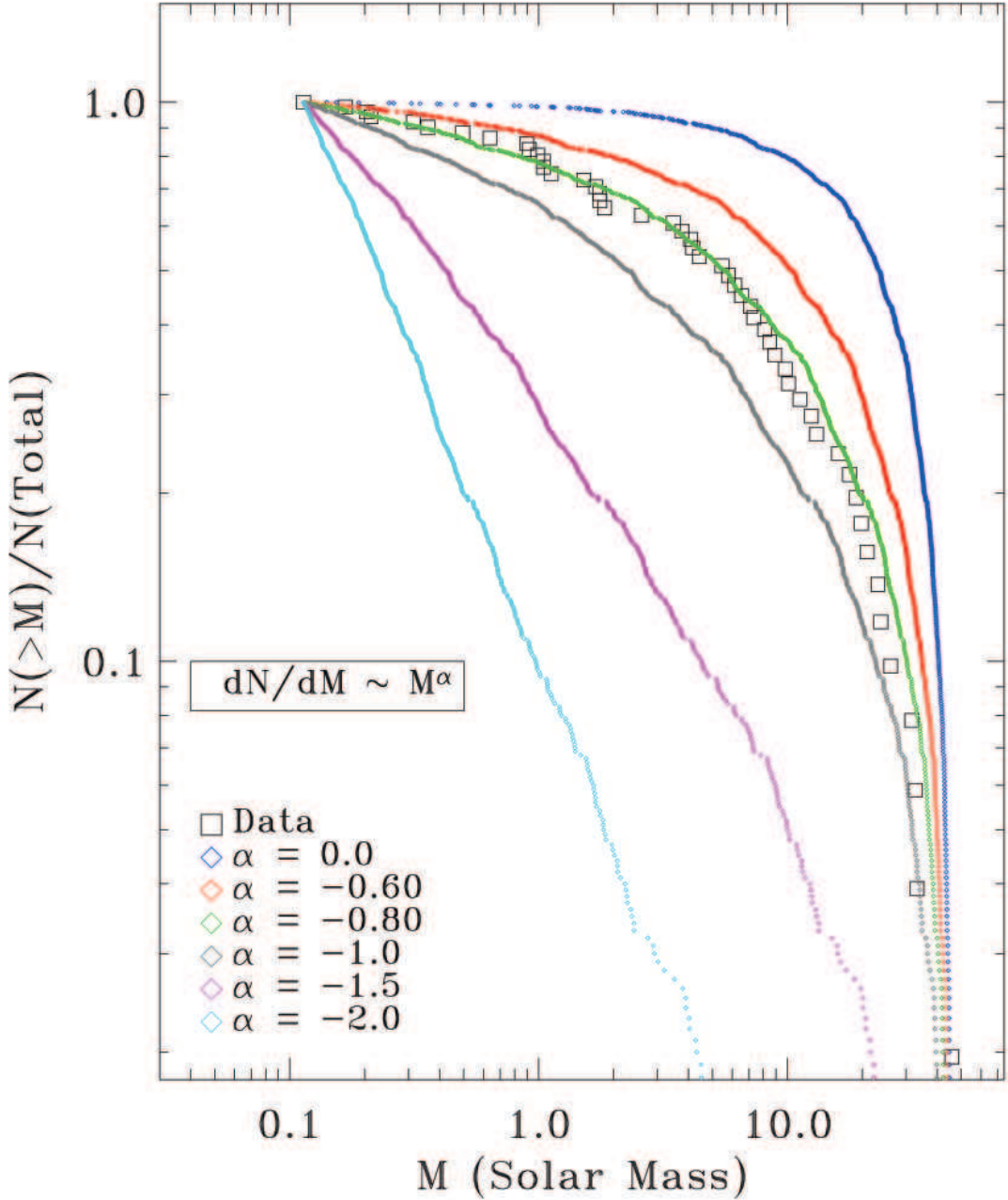}
    \caption{The cumulative CMF of the Orion cores observed in this work are indicated by the open squares.
The cumulative CMFs of simulated core samples based on 5 different single--index underlying power laws mass distributions are plotted as the five colored lines.
For indices flatter than -1.5, the cumulative CMFs show curvature that could  mistakenly be interpreted as indicating two different power laws for the mass distribution at smaller and larger masses.}
      \label{fig:sim}
\end{figure}

As illustrated in Fig.~\ref{fig:sim}, the cumulative mass function does behave as described above. 
Only when $\alpha$ $\leq$  -1.5 does a straight line fit become reasonable.
When the number of cores is large enough to allow binning of the data, the differential core mass function is a straightforward representation of the CMF.
The analysis of such a CMF would give directly the index of a power law mass function and with relatively well--defined statistical uncertainty relatively. 
The advantage of using the cumulative function lies in the fact that the number of data points equals the number of cores, and thus no information is lost. 
A direct power law fit to an arbitrary cumulative function, however, is unreliable. 
If working with the cumulative mass function is unavoidable, a family of models should be generated from a series of power laws with different indices and the best fit model defined as the one that best reproduces the cumulative CMF of the data. 
The caveat for this approach is that the uncertainties are not well defined, as the data points are not independent of each other.

\acknowledgments

This work was supported by the Jet Propulsion Laboratory, California Institute of Technology.  
This work has made use of NASA's Astrophysics Data System. Di Li acknowledges the support coming through the Resident Research Associate
program of the National Research Council and the NASA Postdoctoral Program. 
We thank D. Dowell for his valuable help with observing and data processing. Research at the Caltech Submillimeter Observatory is
supported by NSF grant AST-0229008.

%*********************************************************************************
%Figures
%*********************************************************************************

\newpage
\input{tab1.tex}
\end{document}

%% file: tab1.tex
\begin{deluxetable}{lccccc}
\tablecaption{Quiscent Cores in Orion}
\tablehead{  \colhead{Source Name\tablenotemark{a}}&  \colhead{RA}\tablenotemark{b} & \colhead{DEC \tablenotemark{b,c}}& \colhead{Radius\tablenotemark{d}} & \colhead{Peak Flux}\tablenotemark{e} & \colhead{Mass}\tablenotemark{f}}
\startdata
   ORI1\_1&05 35 26.68&-04 58 12.3&    0.59&     2.6&    0.31\\
   ORI1\_2&05 35 10.76&-04 58 17.1&     1.4&     2.6&     1.8\\
   ORI1\_3&05 35 13.24&-04 59 28.3&     3.4&     5.6&     19.\\
   ORI1\_4&05 35 10.28&-04 59 54.2&     1.9&     2.6&     3.8\\
   ORI1\_5&05 35 15.78&-05 00 01.5&     3.6&     7.9&     26.\\
   ORI1\_6&05 35 17.83&-05 00 18.5&     2.9&     18.&     24.\\
   ORI1\_7&05 35 18.10&-05 00 34.7&     3.2&     11.&     23.\\
   ORI1\_8&05 35 18.80&-05 00 50.9&     2.9&     11.&     21.\\
   ORI1\_9&05 35 20.36&-05 00 52.5&     3.2&     13.&     33.\\
  ORI1\_10&05 35 25.70&-05 01 14.3&     1.0&     2.4&    0.90\\
  ORI1\_11&05 35 22.25&-05 01 14.3&     2.6&     19.&     20.\\
  ORI1\_12&05 35 18.04&-05 01 19.2&     1.0&     2.3&    0.92\\
  ORI1\_13&05 35 23.22&-05 01 29.7&     2.4&     53.&     32.\\
  ORI1\_14&05 35 25.70&-05 01 44.3&    0.73&     2.7&    0.50\\
  ORI1\_15&05 35 17.99&-05 01 44.3&     2.2&     4.3&     6.1\\
  ORI1\_16&05 35 19.99&-05 01 47.5&     2.3&     5.0&     8.1\\
  ORI1\_17&05 35 23.65&-05 02 03.7&     2.1&     6.1&     7.3\\
  ORI1\_18&05 35 25.22&-05 02 38.5&     2.5&     6.8&     9.8\\
   ORI2\_1&05 35 01.31&-05 55 34.5&     1.7&     1.9&     5.4\\
   ORI2\_2&05 35 08.59&-05 55 55.5&     2.9&     7.5&     33.\\
   ORI2\_3&05 35 07.14&-05 56 43.3&     1.2&     2.0&     2.6\\
   ORI2\_4&05 35 11.72&-05 56 57.8&     1.5&     2.1&     4.4\\
   ORI2\_5&05 35 11.89&-05 57 26.2&    0.79&     1.7&     1.0\\
   ORI2\_6&05 35 13.13&-05 57 58.5&     2.1&     4.0&     12.\\
   ORI2\_7&05 35 08.81&-05 58 30.1&     2.0&     3.9&     11.\\
   ORI2\_8&05 35 14.10&-05 58 30.9&    0.79&     1.7&     1.0\\
   ORI4\_1&05 36 04.30&-06 09 40.9&     1.3&    0.92&     1.5\\
   ORI4\_2&05 36 10.66&-06 10 35.1&     2.7&     1.7&     10.\\
   ORI4\_3&05 36 11.42&-06 10 48.1&     2.8&     2.5&     13.\\
   ORI4\_4&05 36 18.05&-06 12 07.4&    0.89&    0.80&    0.64\\
   ORI5\_1&05 36 24.68&-06 13 59.3&     2.6&     1.1&     7.1\\
   ORI5\_2&05 36 25.59&-06 14 12.2&     2.1&     1.1&     4.2\\
   ORI5\_3&05 36 39.78&-06 14 14.7&    0.74&    0.63&    0.36\\
   ORI5\_4&05 36 36.71&-06 14 59.2&     2.6&     2.0&     8.9\\
   ORI5\_5&05 36 37.73&-06 15 39.6&    0.58&    0.62&    0.21\\
   ORI7\_1&05 35 56.66&-06 25 51.2&     2.6&    0.34&     3.5\\
   ORI7\_2&05 35 47.43&-06 26 17.9&     1.9&    0.33&     1.8\\
   ORI7\_3&05 35 52.83&-06 27 43.7&     2.0&    0.30&     1.8\\
 ORI8nw\_1&05 36 20.92&-06 21 58.4&     1.9&     2.8&     4.1\\
 ORI8nw\_2&05 36 18.60&-06 22 19.4&     3.4&     19.&     46.\\
 ORI8nw\_3&05 36 23.67&-06 24 52.4&     1.4&     1.9&     1.7\\
 ORI8se\_1&05 36 24.48&-06 22 41.3&     2.6&     3.1&     8.5\\
 ORI8se\_2&05 36 25.45&-06 22 48.6&     2.6&     2.0&     6.5\\
 ORI8se\_3&05 36 25.83&-06 23 20.1&    0.58&    0.73&    0.17\\
 ORI8se\_4&05 36 23.67&-06 23 30.6&     1.3&    0.93&    0.99\\
 ORI8se\_5&05 36 25.18&-06 24 56.4&     3.8&     2.0&     18.\\
 ORI8se\_6&05 36 31.44&-06 25 23.1&    0.48&    0.73&    0.11\\
 ORI8se\_7&05 36 41.75&-06 26 17.3&     3.9&     2.0&     16.\\
 ORI8se\_8&05 36 32.41&-06 26 19.8&     1.4&    0.80&     1.1\\
 ORI8se\_9&05 36 30.47&-06 26 57.0&    0.64&    0.72&    0.20\\
  ORI11\_1&05 36 16.89&-06 38 04.2&     1.9&     1.2&     5.8\\
\enddata
\tablenotetext{a}{The source is identified by the survey region where the core is located followed by a sub number assigned according to its declination.}
\tablenotetext{b}{The RA and DEC are presented in the format of HHMMSS.SS and DDMMSS.S (J2000).}
\tablenotetext{c}{The sources are sorted from north to south.}
\tablenotetext{d}{The mean value of the semi-major and semi-minor axes in the units of 0.01 pc}
\tablenotetext{e}{In the units of Jy per 9\arcsec\ beam}
\tablenotetext{f}{Solar masses}
\end{deluxetable}